\documentclass[11pt]{article}

\usepackage{amsfonts}
\usepackage{amsmath}
\usepackage{graphicx}
\usepackage{subfigure}
\usepackage{color}
\usepackage{vmargin}

\setpapersize{USletter}
\setmarginsrb{1in}{0.75in}{1in}{1.25in}{0in}{0in}{11pt}{0.4in} 
\newlength{\defbaselineskip}
\begin{document}

\title{
Algorithmic and Statistical Perspectives \\ 
on Large-Scale Data Analysis\footnote{To 
appear in Uwe Naumann and Olaf Schenk, editors, \emph{Combinatorial 
Scientific Computing}, Chapman and Hall/CRC Press, 2011.} 
}

\author{
Michael~W.~Mahoney%
\thanks{
Department of Mathematics, 
Stanford University, 
Stanford, CA 94305. 
Email: mmahoney@cs.stanford.edu
}
}

\date{}
\maketitle


\begin{abstract}
\noindent
In recent years, ideas from statistics and scientific computing have begun
to interact in increasingly sophisticated and fruitful ways with ideas from 
computer science and the theory of algorithms to aid in the development of 
improved worst-case algorithms that are useful for large-scale scientific 
and Internet data analysis problems.  
In this chapter, I will describe two recent examples---one having to do with 
selecting good columns or features from a (DNA Single Nucleotide 
Polymorphism) data matrix, and the other having to do with selecting good 
clusters or communities from a data graph (representing a social or 
information network)---that drew on ideas from both areas and that
may serve as a model for exploiting complementary algorithmic and statistical
perspectives in order to solve applied large-scale data analysis problems.
\end{abstract}

\section{Introduction}

In recent years, motivated in large part by technological advances in both
scientific and Internet domains, there has been a great deal of interest in 
what may be broadly termed \emph{large-scale data analysis}.
In this chapter, I will focus on what seems to me to be a remarkable and
inevitable trend in this increasingly-important area.
This trend has to do with the convergence of two very different perspectives 
or worldviews on what are appropriate or interesting or fruitful ways to 
view the data.
At the risk of oversimplifying a large body of diverse work, I would like to 
draw a distinction between what I will loosely term the \emph{algorithmic 
perspective} and the \emph{statistical perspective} on large-scale data 
analysis problems.
By the former, I mean roughly the approach that one trained in computer 
science might adopt.
From this perspective, primary concerns include database issues, algorithmic 
questions such as models of data access, and the worst-case running time of 
algorithms for a given objective function.
By the latter, I mean roughly the approach that one trained in statistics 
(or some application area where strong domain-specific assumptions about the 
data may be made) might adopt.
From this perspective, primary concerns include questions such as how well 
the objective functions being considered conform to the phenomenon under 
study and whether one can make reliable predictions about the world from 
the data at hand.

Although very different, these two approaches are certainly not incompatible.
Moreover, in spite of peoples' best efforts \emph{not} to forge the union 
between these two worldviews, and instead to view the data from one 
perspective or another, depending on how one happened to be trained, the 
large-scale data analysis problems that we are increasingly facing---in 
scientific, Internet, financial, etc.  applications---are so important and 
so compelling---either from a business perspective, or an academic 
perspective, or an intellectual perspective, or a national security 
perspective, or from whatever perspective you find to be most 
compelling---that we are being forced to forge this union.

Thus, \emph{e.g.}, if one looks at the SIG-KDD meeting, ACM's flagship 
meeting on data analysis, now (in $2010$) versus $10$ or $15$ years ago, one 
sees a substantial shift away from more database-type issues toward topics 
that may be broadly termed statistical, \emph{e.g.}, that either involve 
explicit statistical modeling or involve making generative assumptions about 
data elements or network participants.
And vice-versa---there has been a substantial shift in statistics, and 
especially in the natural and social sciences more generally, toward 
thinking in great detail about computational issues.
Moreover, to the extent that statistics as an academic discipline is 
deferring on this matter, the relatively new area of machine learning is 
filling in the gap.

I should note that I personally began thinking about these issues some time 
ago. 
My Ph.D. was in computational statistical mechanics, and it involved a lot 
of Monte Carlo and molecular dynamics computations on liquid water and 
DNA-protein-water interactions.
After my dissertation, I switched fields to theoretical computer science, 
where I did a lot of work on the theory (and then later on the application) 
of randomized algorithms for large-scale matrix problems.
One of the things that struck me during this transition was the deep 
conceptual disconnect between these two areas.
In computer science, we have a remarkable infrastructure of 
machinery---from complexity classes and data structuring and algorithmic 
models, to database management, computer architecture, and software 
engineering paradigms---for solving problems.
On the other hand, it seems to me that there tends to be a remarkable 
lack of appreciation, and thus associated cavalierness, when it comes to 
understanding how the data can be messy and noisy and poorly-structured in ways that 
adversely affect how one can be confident in the conclusions that one draws 
about the world as a result of the output of one's fast algorithms.

In this chapter, I would like to describe some of the fruits of this 
thought.
To do so, I will focus on two very particular very applied problems in 
large-scale data analysis on which I have worked.
The solution to these two problems benefited from taking advantage of 
the complementary algorithmic versus statistical perspectives in a novel 
way.
In particular, we will see that, by understanding the statistical properties 
\emph{implicit} in worst-case algorithms, we can make very strong claims 
about very applied problems.
These claims would have been \emph{much} more difficult to make and justify 
if one chose to view the problem from just one perspective or the other.
The first problem has to do with selecting good features from a data matrix 
representing DNA microarray or DNA Single Nucleotide Polymorphism 
data.
This is of interest, \emph{e.g.}, to geneticists who want to understand 
historical trends in population genetics, as well as to biomedical 
researchers interested in so-called personalized medicine.
The second problem has to do with identifying, or certifying that there do 
not exist, good clusters or communities in a data graph representing a 
large social or information network.
This is of interest in a wide range of applications, such as finding good 
clusters or micro-markets for query expansion or bucket testing in Internet 
advertising applications.

The applied results for these two problems have been reported previously in 
appropriate domain-specific (in this case, genetics and Internet)
publications~\cite{Paschou07a,Paschou07b,CUR_PNAS,LLDM08_communities_CONF,LLDM08_communities_TR,LLM10_communities_CONF},
and I am indebted to my collaborators with whom I have discussed some of
these ideas in preliminary form. 
Thus, rather than focusing on the genetic issues \emph{per se} or the 
Internet advertising issues \emph{per se} or the theoretical analysis 
\emph{per se}, in this chapter I would like to focus on what was going on 
``under the hood'' in terms of the interplay between the algorithmic and 
statistical perspectives.
The hope is that these two examples can serve as a model for exploiting the 
complementary aspects of the algorithmic and statistical perspectives in 
order to solve very applied large-scale data analysis problems more 
generally.%
\footnote{This chapter grew out of an invited talk at a seminar at Schloss 
Dagstuhl on Combinatorial Scientific Computing (a research area that sits at 
the interface between algorithmic computer science and traditional 
scientific computing).
While the connection between ``combinatorial'' and ``algorithmic'' in my 
title is hopefully obvious, the reader should also note a connection between 
the statistical perspective I have described and approaches common in 
traditional scientific computing.
For example, even if a formal statistical model is not specified, the fact 
that hydrated protein-DNA simulations or fluid dynamics simulations ``live'' 
in two or three dimensions places very strong regularity constraints on the 
ways in which information can propagate from point to point and thus on the 
data sets derived.
Moreover, in these applications, there is typically a great deal of 
domain-specific knowledge that can be brought to bear that informs the 
questions one asks, and thus one clearly observes a tension between 
worst-case analysis versus field-specific assumptions.
This implicit knowledge should be contrasted with the knowledge available 
for matrices and graphs that arise in much more unstructured data analysis 
applications such as arise in certain genetics and in many Internet 
domains.}  
As we will see, in neither of these two applications did we \emph{first} 
perform statistical modeling, independent of algorithmic considerations, and 
\emph{then} apply a computational procedure as a black box.
This approach of more closely coupling the computational procedures used 
with statistical modeling or statistical understanding of the data seems 
particularly appropriate more generally for very large-scale data analysis 
problems, where design decisions are often made based on computational 
constraints but where it is of interest to understand the implicit 
statistical consequences of those decisions.

\section{Diverse approaches to modern data analysis problems}

Before proceeding, I would like in this section to set the stage by 
describing in somewhat more detail some of diverse approaches that have been 
brought to bear on modern data analysis 
problems~\cite{FSS96,Smy00,Donoho00,Bri01_all,Lam03,PS03,MMDS06,MMDS08_SiamNews}.
In particular, although the significant differences between the algorithmic 
perspective and the statistical perspective have been highlighted 
previously~\cite{Lam03}, they are worth reemphasizing.


A common view of the data in a database, in particular historically among 
computer scientists interested in data mining and knowledge discovery, has 
been that the data are an accounting or a record of everything that 
happened in a particular setting.
For example, the database might consist of all the customer transactions 
over the course of a month, or it might consist of all the friendship 
links among members of a social networking site.
From this perspective, the goal is to tabulate and process the data at hand 
to find interesting patterns, rules, and associations.
An example of an association rule is the proverbial ``People who buy beer
between $5$ p.m.\ and $7$ p.m.\ also buy diapers at the same time.''
The performance or quality of such a rule is judged by the fraction of the 
database that satisfies the rule exactly, which then boils down to the 
problem of finding frequent itemsets.
This is a computationally hard problem, and much algorithmic work has been 
devoted to its exact or approximate solution under different models of data 
access.

A very different view of the data, more common among statisticians, is one of 
a particular random instantiation of an underlying process describing 
unobserved patterns in the world.
In this case, the goal is to extract information about the world from the 
noisy or uncertain data that are observed.
To achieve this, one might posit a model: 
$ \mathit{data} \sim F_{\theta} $ and 
$ \operatorname*{mean}(\mathit{data}) = g(\theta) $,
where $F_{\theta}$ is a distribution that describes the random variability
of the data around the deterministic model $g(\theta)$ of the data.
Then, using this model, one would proceed to analyze the data to make 
inferences about the underlying processes and predictions about future 
observations. 
From this perspective, modeling the noise component or variability well is as 
important as modeling the mean structure well, in large part since 
understanding the former is necessary for understanding the quality of 
predictions made.
With this approach, one can even make predictions about events that have 
yet to be observed.
For example, one can assign a probability to the event that a given user 
at a given web site will click on a given advertisement presented at a 
given time of the day, even if this particular event does not exist in the 
database.

Although these two perspectives are certainly not incompatible, they are 
very different, and they lead one to ask very different questions of the 
data and of the structures that are used to model data.
Recall that in many applications, graphs and matrices are common ways to 
model the data~\cite{MMDS06,MMDS08_SiamNews}.
For example, a common way to model a large social or information network is 
with an interaction graph model, $G=(V,E)$, in which nodes in the vertex set 
$V$ represent ``entities'' and the edges (whether directed, undirected, 
weighted or unweighted) in the edge set $E$  represent ``interactions'' 
between pairs of entities. 
Alternatively, these and other data sets can be modeled as 
matrices, since an $m \times n$ real-valued matrix $A$ provides a natural 
structure for encoding information about $m$ objects, each of which is 
described by $n$ features.  
Thus, in the next two sections, I will describe two recent examples---one
having to do with modeling data as matrices, and the other having to do with
modeling data as a graph---of particular data analysis problems that 
benefited from taking advantage in novel ways of the respective strengths 
of the algorithmic and statistical perspectives.

\section{Genetics applications and novel matrix algorithms}

In this section, I will describe an algorithm for selecting a ``good'' set of 
exactly $k$ columns (or, equivalently, exactly $k$ rows) from an arbitrary 
$m \times n$ matrix $A$.  
This problem has been studied extensively in scientific computing and 
Numerical Linear Algebra (NLA), often motivated by the goal of finding a 
good basis with which to perform large-scale numerical computations.
In addition, variants of this problem have recently received a great deal of 
attention in Theoretical Computer Science (TCS).  
More generally, problems of this sort arise in many data analysis 
applications, often in the context of finding a good set of features with 
which to describe the data or to perform tasks such as classification or 
regression.

\subsection{Motivating genetics application}

Recall that ``the human genome'' consists of a sequence of roughly $3$ 
billion base pairs on $23$ pairs of chromosomes, roughly $1.5\%$ of which 
codes for approximately $20,000$ -- $25,000$ proteins.
A DNA microarray is a device that can be used to measure simultaneously the 
genome-wide response of the protein product of each of these genes for an 
individual or group of individuals in numerous different environmental 
conditions or disease states.
This very coarse measure can, of course, hide the individual differences or 
polymorphic variations. 
There are numerous types of polymorphic variation, but the most amenable to 
large-scale applications is the analysis of Single Nucleotide Polymorphisms 
(SNPs), which are known locations in the human genome where two alternate 
nucleotide bases (or alleles, out of $A$, $C$, $G$, and $T$) are observed in 
a non-negligible fraction of the population.
These SNPs occur quite frequently, ca. $1$ b.p. per thousand, and thus they
are effective genomic markers for the tracking of disease genes (\emph{i.e.}, 
they can be used to perform classification into sick and not sick) as well as
population histories (\emph{i.e}, they can be used to infer properties about 
population genetics and human evolutionary history).

In both cases, $m \times n$ matrices $A$ naturally arise, either as a
people-by-gene matrix, in which $A_{ij}$ encodes information about the 
response of the $j^{th}$ gene in the $i^{th}$ individual/condition, or as 
people-by-SNP matrices, in which $A_{ij}$ encodes information about the 
value of the $j^{th}$ SNP in the $i^{th}$ individual.
Thus, matrix computations have received attention in these 
applications~\cite{Alter_SVD_00,KPS02,Meng03,Horne04,LA04}.
A common \emph{modus operandi} in applying NLA and matrix techniques such 
as PCA and the SVD to to DNA microarray, DNA SNPs, and other data problems 
is:
\begin{itemize}
\item
Model the people-by-gene or people-by-SNP data as an $m \times n$ matrix $A$.
\item
Perform the SVD (or related eigen-methods such as PCA or recently-popular 
manifold-based methods~\cite{TSL00,RS00,SWHSL06} that boil down to the SVD) 
to compute a small number of eigengenes or eigenSNPs or eigenpeople that 
capture most of the information in the data matrix.
\item 
Interpret the top eigenvectors as meaningful in terms of underlying 
biological processes; or apply a heuristic to obtain actual genes or actual 
SNPs from the corresponding eigenvectors in order to obtain such an 
interpretation.
\end{itemize}
In certain cases, such reification may lead to insight and such heuristics 
may be justified. 
(For instance, if the data happen to be drawn from a Guassian distribution, 
then the eigendirections tend to correspond to the axes of the corresponding
ellipsoid, and there are many vectors that, up to noise, point along those
directions.)
In such cases, however, the justification comes from domain knowledge and 
not the mathematics~\cite{Gould96,KPS02,CUR_PNAS}.
The reason is that the eigenvectors themselves, being mathematically defined
abstractions, can be calculated for any data matrix and thus are not easily 
understandable in terms of processes generating the data:
eigenSNPs (being linear combinations of SNPs) cannot be assayed;
nor can eigengenes be isolated and purified;
nor is one typically interested in how eigenpatients respond to treatment 
when one visits a physician.

For this and other reasons, a common task in genetics and other areas of data 
analysis is the following: given an input data matrix $A$ and a parameter 
$k$, find the best subset of exactly $k$ \emph{actual} DNA SNPs or 
\emph{actual} genes, \emph{i.e.}, \emph{actual} columns or rows from $A$, to 
use to cluster individuals, reconstruct biochemical pathways, reconstruct 
signal, perform classification or inference, etc.

\subsection{A formalization of and prior approaches to this problem}

Unfortunately, common formalizations of this algorithmic problem---including 
looking for the $k$ actual columns that capture the largest amount of 
information or variance in the data or that are maximally 
uncorrelated---lead to intractable optimization 
problems~\cite{CM09a,CM09b}.
In this chapter, I will consider the so-called Column Subset Selection 
Problem (CSSP):
given as input an arbitrary $m \times n$ matrix $A$ and a rank parameter 
$k$, choose the set of exactly $k$ columns of $A$ s.t. the $m \times k$ 
matrix $C$ minimizes (over all ${n \choose k}$ sets of such columns) the error:
\begin{equation}
\min ||A-P_CA||_{\xi} = \min ||A-CC^+A||_{\xi}  \hspace{5mm} (\xi=2,F)
\label{eqn:error-measure}
\end{equation}
where  $\xi=2,F$ represents the spectral or Frobenius norm%
\footnote{Recall that the spectral norm is the largest singular value of the 
matrix, and thus it is a ``worst case'' norm in that it measures the 
worst-case stretch of the matrix, while the Frobenius norm is more of an 
``averaging'' norm, since it involves a sum over every singular direction.
The former is of greater interest in scientific computing and NLA, where one 
is interested in actual columns for the subspaces they define and for their 
good numerical properties, while the latter is of greater interest in data 
analysis and machine learning, where one is more interested in actual 
columns for the features they define.}
of $A$ and where $P_C=CC^+$ is the projection onto the subspace spanned by 
the columns of~$C$.

Within NLA, a great deal of work has focused on this CSSP
problem~\cite{BG65,Fos86,Cha87,CH90,BH91,HP92,CI94,GE96,BQ98a,PT99,Pan00}.
Several general observations about the NLA approach include:
\begin{itemize}
\item
The focus in NLA is on \emph{deterministic algorithms}.
Moreover, these algorithms are greedy, in that at each iterative step, the 
algorithm makes a decision about which columns to keep according to a 
pivot-rule that depends on the columns it currently has, the spectrum of 
those columns, etc.
Differences between different algorithms often boil down to how deal with 
such pivot rules decisions, and the hope is that more sophisticated 
pivot-rule decisions lead to better algorithms in theory or in practice.
\item
There are deep \emph{connections with QR factorizations} and in particular 
with the so-called Rank Revealing QR factorizations.
Moreover, there is an emphasis on optimal conditioning questions, backward 
error analysis issues, and whether the running time is a large or small 
constant multiplied by $n^2$ or $n^3$.
\item
Good \emph{spectral norm bounds} are obtained.
A typical spectral norm bound is:
$$
||A-P_CA||_2 \le O\left(\sqrt{k(n-k)}\right)||A-P_{U_k}A||_2 ,
$$
and these results are algorithmic, in that the running time is a low-degree 
polynomial in $m$ and $n$~\cite{GE96}.
On the other hand, the strongest results for the Frobenius norm in this 
literature is
$$
||A-P_CA||_F \le \sqrt{(k+1)(n-k)}||A-P_{U_k}A||_2  ,
$$
but it is only an existential result, \emph{i.e.}, the only known algorithm 
essentially involves exhaustive enumeration~\cite{HP92}.
(In these two expressions, $U_k$ is the $m \times k$ matrix consisting of the
top $k$ left singular vectors of $A$, and $P_{U_k}$ is a projection matrix 
onto the span of $U_k$.)
\end{itemize}

Within TCS, a great deal of work has focused on the related problem of 
choosing good columns from a 
matrix~\cite{DFKVV04_JRNL,dkm_matrix1,dkm_matrix2,dkm_matrix3,RV07,DMM08_CURtheory_JRNL}.
Several general observations about the TCS approach include:
\begin{itemize}
\item
The focus in TCS is on \emph{randomized 
algorithms}.
In particular, with these algorithms, there exists some nonzero 
probability, which can typically be made extremely small, say $10^{-20}$, 
that the algorithm will return columns that fail to satisfy the desired 
quality-of-approximation bound. 
\item
The algorithms select \emph{more than $k$ columns}, and the best 
rank-$k$ projection onto those columns is considered.
The number of columns is typically a low-degree polynomial in $k$, most often
$O(k \log k)$, where the constants hidden in the big-O notation are quite 
reasonable.
\item
Very good \emph{Frobenius norm bounds} are obtained. 
For example, the algorithm (described below) that provides the strongest 
Frobenius norm bound achieves:
\begin{equation}
||A-P_{C_k}A||_F \le (1+\epsilon) ||A-P_{U_k}A ||_F  ,
\label{eqn:rel-err}
\end{equation}
while running in time of the order of computing an exact or approximate 
basis for the top-$k$ right singular subspace~\cite{DMM08_CURtheory_JRNL}.
The TCS literature also demonstrates that there exists a set of $k$ columns 
that achieves a constant-factor approximation:
$$
||A-P_{C_k}A||_F \le \sqrt{k} ||A-P_{U_k}A ||_F  ,
$$
but note that this is an existential result~\cite{DV06}.
(Here, $C_k$ is the best rank-$k$ approximation to the matrix $C$, and 
$P_{C_k}$ is the projection matrix onto this $k$-dimensional space.)
\end{itemize}
Much of the early work in TCS focused on randomly sampling columns according 
to an importance sampling distribution that depended on the Euclidean norm 
of those columns~\cite{DFKVV04_JRNL,dkm_matrix1,dkm_matrix2,dkm_matrix3,RV07}.
This had the advantage of being ``fast,'' in the sense that it could be
performed in a small number of ``passes'' over that data from external 
storage, and also that additive-error quality-of-approximation bounds could 
be proved.
This had the disadvantage of being less immediately-applicable to scientific 
computing and large-scale data analysis applications.
For example, columns are often normalized during data preprocessing; and 
even when not normalized, column norms can still be uninformative, as in 
heavy-tailed graph data%
\footnote{By \emph{heavy-tailed graph}, I mean a graph (or equivalently the
adjacency matrix of such a graph) such as the social and information networks
described below, in which quantities such as the degree distribution or 
eigenvalue distribution decay in a heavy-tailed or power law manner.}
 where they often correlate strongly with 
simpler statistics such as node degree.

The algorithm from the TCS literature that achieves the strongest Frobenius 
norm bounds of the form~(\ref{eqn:rel-err}) is the following.%
\footnote{Given a matrix $A$ and a rank parameter $k$, one can express the 
SVD as $A=U \Sigma V^T$, in which case the best rank-$k$ approximation to 
$A$ can be expressed as $A_k = U_k \Sigma_k V_k^T$.  In the text, I will 
sometimes overload notation and use $V_k^T$ to refer to any $k \times n$ 
orthonormal matrix spanning
the space spanned by the top-$k$ right singular vectors.  The reason is that
this basis is used only to compute the importance sampling probabilities; 
since those probabilities are proportional to the diagonal elements of the 
projection matrix onto the span of this basis, the particular basis does not
matter.} 
Given an $m \times n$ matrix $A$ and a rank parameter $k$:
\begin{itemize}
\item
Compute the importance sampling probabilities $\{p_i\}_{i=1}^{n}$, where 
$p_i=\frac{1}{k}||{V_k^T}^{(i)}||_2^2$, where $V_k^T$ is \emph{any} 
$k \times n$ orthogonal matrix spanning the top-$k$ right singular subspace 
of~$A$.
(Note that these quantities are proportional to the diagonal elements of the
projection matrix onto the span of $V_k^T$.)
\item
Randomly select and rescale $c = O(k \log k /\epsilon^2)$ columns of $A$
according to these probabilities.
\end{itemize}
A more detailed description of this algorithm may be found 
in~\cite{DMM08_CURtheory_JRNL,CUR_PNAS},
where it is proven that~(\ref{eqn:rel-err}) 
holds with extremely high probability.
The computational bottleneck for this algorithm is computing 
$\{p_i\}_{i=1}^{n}$, for which it suffices to compute \emph{any} 
$k \times n$ matrix $V_k^T$ that spans the top-$k$ right singular subspace 
of $A$.
(That is, it suffices to compute any orthonormal basis spanning $V_k^T$, and
it is not necessary to compute the full SVD.)
It is an open problem whether these importance sampling probabilities can 
be approximated more rapidly. 

To motivate the importance sampling probabilities used by this algorithm, 
recall that if one is looking for a worst-case relative-error approximation 
of the form~(\ref{eqn:rel-err}) to a matrix with $k-1$ large singular values 
and one much smaller singular value, then the directional information of the 
$k^{th}$ singular direction will be hidden from the Euclidean norms of the 
matrix.
Intuitively, the reason is that, since $A_k=U_k \Sigma_k V_k^T$, the 
Euclidean norms of the columns of $A$ are convolutions of ``subspace 
information'' (encoded in $U_k$ and $V_k^T$) and ``size-of-$A$ information'' 
(encoded in $\Sigma_k$).
This suggests deconvoluting subspace information and size-of-$A$ 
information by choosing importance sampling probabilities that depend on the
Euclidean norms of the columns of $V_k^T$.
Thus, this importance sampling distribution defines a nonuniformity 
structure over $\mathbb{R}^{n}$ that indicates \emph{where} in the 
$n$-dimensional space the information in $A$ is being sent, independent of 
\emph{what} that (singular value) information is.
As we will see in the next two subsections, by using these importance 
sampling probabilities, we can obtain novel algorithms for two very 
traditional problems in NLA and scientific computing.

\subsection{An aside on least squares and statistical leverage}
\label{sxn:LS}

The analysis of the relative-error algorithm described in the previous 
subsection and of the algorithm for the CSSP described in the next 
subsection boils down to a least-squares approximation result.
Intuitively, these algorithms find columns that provide a space that is 
good in a least-squares sense, when compared to the best rank-$k$ space, at 
reconstructing every row of the input
matrix~\cite{DMM08_CURtheory_JRNL,BMD08_CSSP_TR}.

Thus, consider the problem of finding a vector $x$ such that $Ax \approx b$, 
where the rows of $A$ and elements of $b$ correspond to constraints and the 
columns of $A$ and elements of $x$ correspond to variables.
In the very overconstrained case where the $m \times n$ matrix $A$ has 
$m \gg n$,%
\footnote{In this section only, we will assume that $m \gg n$.}
there is in general no vector $x$ such that $Ax=b$, and it is 
common to quantify ``best'' by looking for a vector $x_{opt}$ such that the 
Euclidean norm of the residual error is small, \emph{i.e.}, to solve 
the least-squares (LS) approximation problem
$$
x_{opt} = \mbox{argmin}_x ||Ax-b||_2  .
$$
This problem is ubiquitous in applications, where it often arises from 
fitting the parameters of a model to experimental data, and it is central 
to theory.
Moreover, it has a natural statistical interpretation as providing the best 
estimator within a natural class of estimators, and it has a natural 
geometric interpretation as fitting the part of the vector $b$ that resides 
in the column space of $A$.
From the viewpoint of low-rank matrix approximation and the CSSP, this LS 
problem arises since measuring the error with a Frobenius or spectral 
norm, as in~(\ref{eqn:error-measure}), amounts to choosing columns that are 
``good'' in a least squares sense.

From an algorithmic perspective, the relevant question is: how long does it 
take to compute $x_{opt}$?
The answer here is that is takes $O(mn^2)$ time~\cite{GVL96}---\emph{e.g.}, 
depending on numerical issues, condition numbers, etc., this can be 
accomplished with the Cholesky decomposition, a variant of the QR 
decomposition, or by computing the full SVD. 

From a statistical perspective, the relevant question is: when is computing 
this $x_{opt}$ the right thing to do?
The answer to this is that this LS optimization is the right problem to 
solve when the relationship between the ``outcomes'' and ``predictors'' is 
roughly linear and when the error processes generating the data are ``nice'' 
(in the sense that they have mean zero, constant variance, are uncorrelated, 
and are normally distributed; or when we have adequate sample size to rely 
on large sample theory)~\cite{ChatterjeeHadi88}.

Of course, in practice these assumptions do not hold perfectly, and a 
prudent data analyst will check to make sure that these assumptions have 
not been too violated.
To do this, it is common to assume that $b=Ax+\varepsilon$, where $b$ is the 
response, the columns $A^{(i)}$ are the carriers, and $\varepsilon$ is the 
nice error process.
Then $x_{opt}=(A^TA)^{-1}A^Tb$, and thus $\hat{b}=Hb$, where the projection 
matrix onto the column space of $A$, $H=A(A^TA)^{-1}A^T$, is the so-called 
\emph{hat matrix}.
It is known that $H_{ij}$ measures the influence or statistical leverage 
exerted on the prediction $\hat{b}_i$ by the observation 
$b_j$~\cite{HW78,ChatterjeeHadi88,CH86}.
Relatedly, if the $i^{th}$ diagonal element of $H$ is particularly large 
then the $i^{th}$ data point is particularly sensitive or influential in 
determining the best LS fit, thus justifying the interpretation of the 
elements $H_{ii}$ as \emph{statistical leverage scores}~\cite{CUR_PNAS}.%
\footnote{This concept has also proven useful under the name of graph 
resistance~\cite{SS08a_STOC} and also of coherence~\cite{CR08_TR}.}

\begin{figure}
   \begin{center}
      \subfigure[Wood Beam Data]{
         \includegraphics[width=0.45\textwidth]{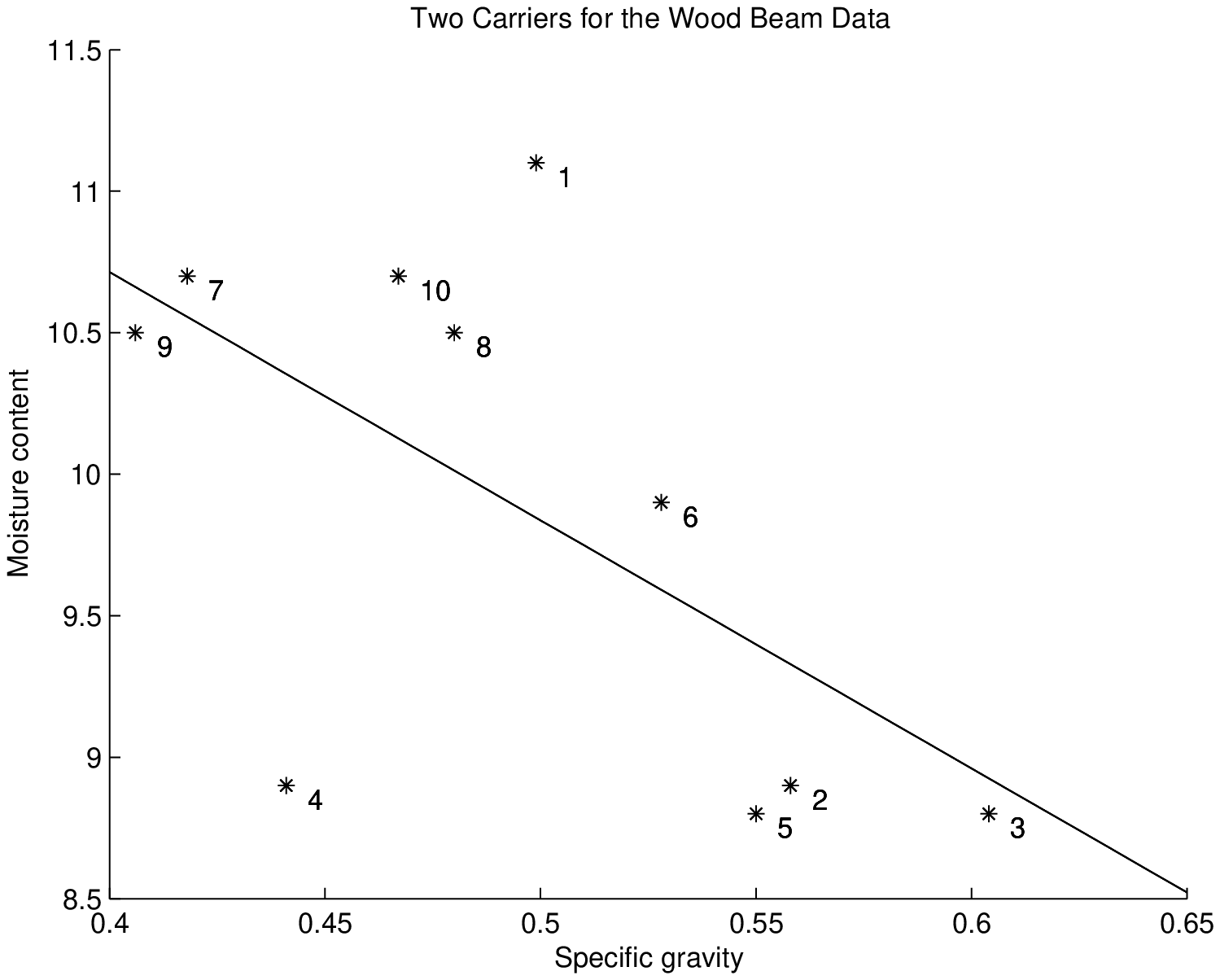}
         \label{fig:leverage:wooddata}
      } \qquad 
      \subfigure[Corresponding Leverage Scores]{
         \hspace{20mm}
         \includegraphics[width=0.08\textwidth]{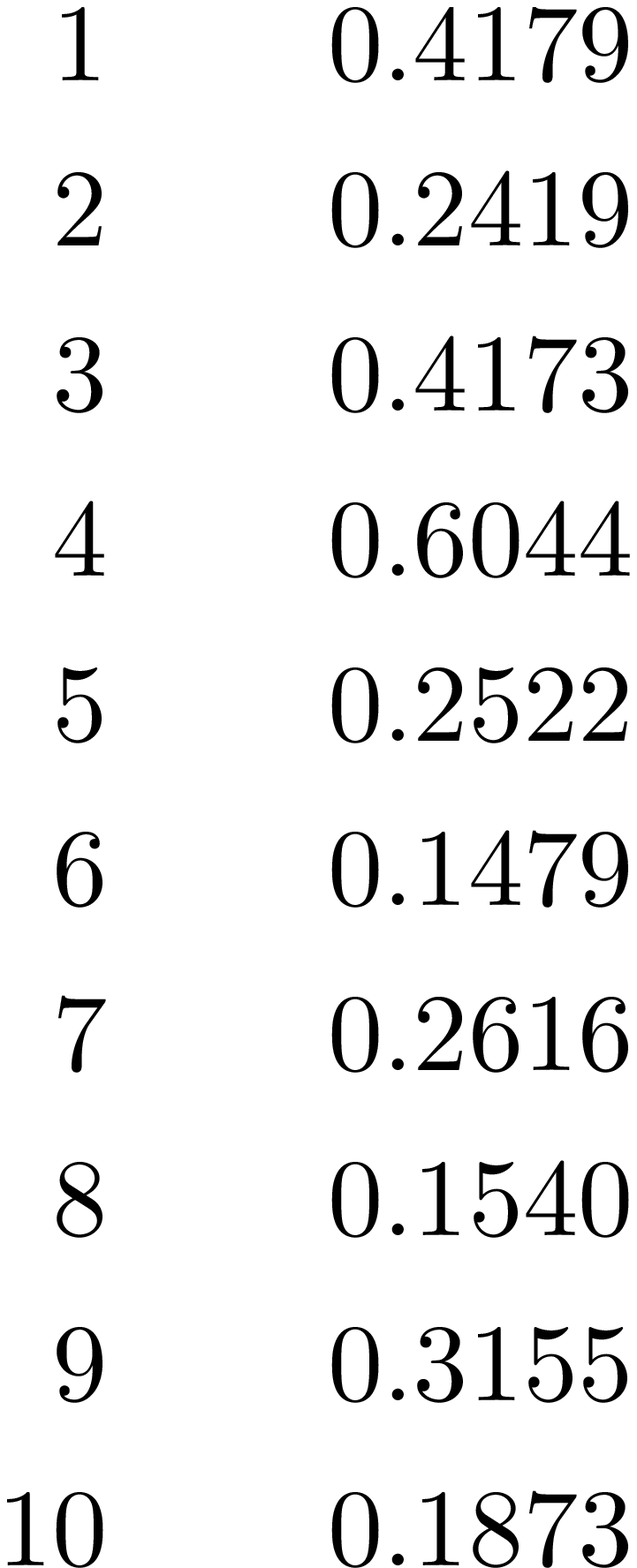}
         \hspace{20mm}
         \label{fig:leverage:woodscores}
      } \\
      \subfigure[Zachary Karate Club Data]{
         \includegraphics[width=0.40\textwidth]{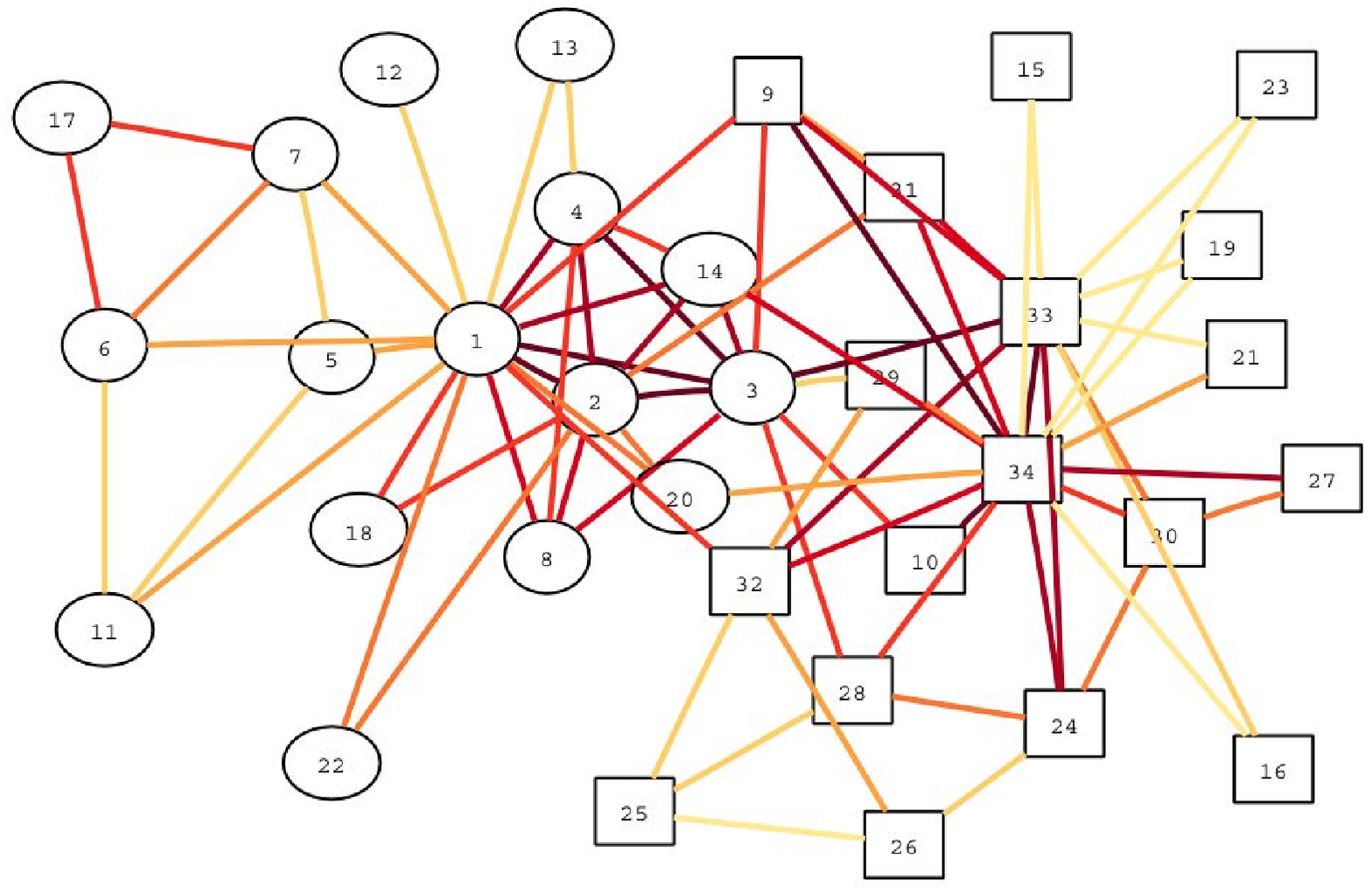}
         \label{fig:leverage:zachary}
      } \qquad 
      \subfigure[Cumulative Leverage]{
         \includegraphics[width=0.45\textwidth]{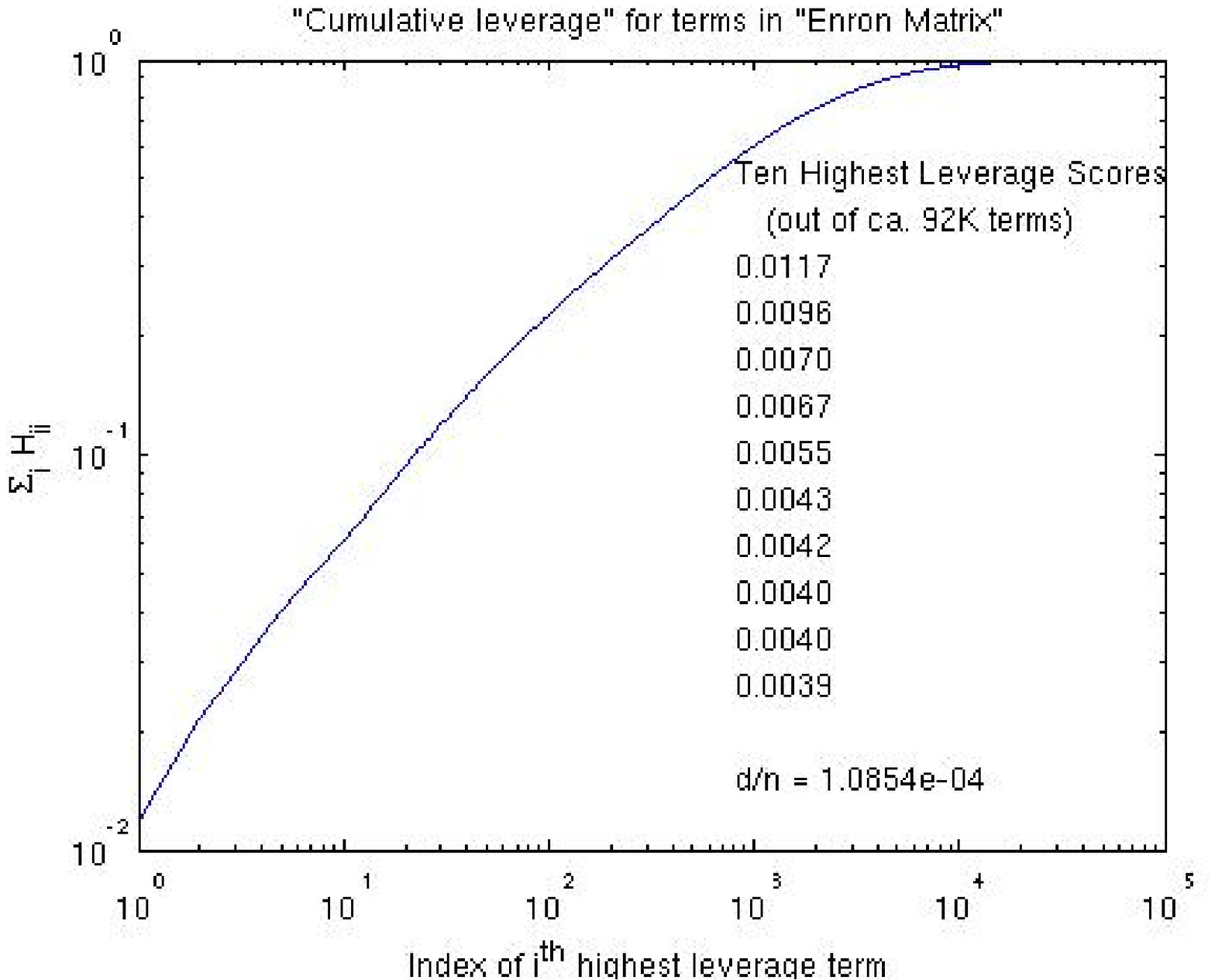}
         \label{fig:leverage:cumlev}
      } \\
      \subfigure[Leverage Score and Information Gain for DNA Microarray Data]{
         \includegraphics[width=1.00\textwidth]{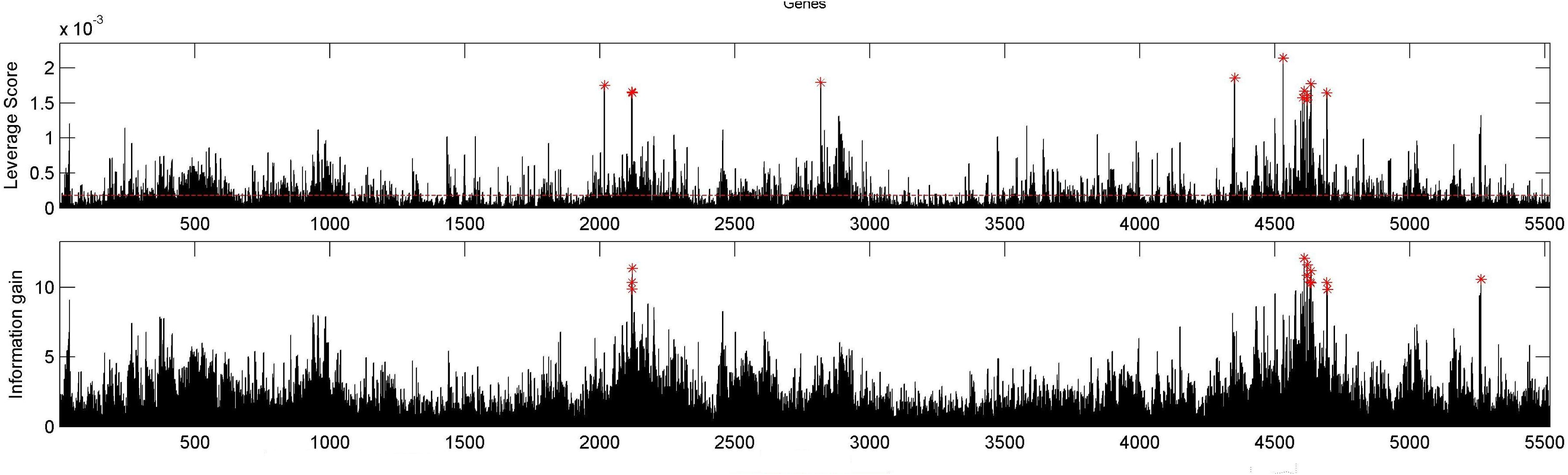}
         \label{fig:leverage:bio}
      } 
   \end{center}
\vspace{-5mm}
\caption{
(\ref{fig:leverage:wooddata})
The Wood Beam Data described in~\cite{HW78} is an example illustrating the
use of statistical leverage scores in the context of least-squares 
approximation.
Shown are the original data and the best least-squares fit.
(\ref{fig:leverage:woodscores})
The leverage scores for each of the ten data points in the Wood Beam Data.
Note that the data point marked ``4'' has the largest leverage score, as 
might be expected from visual 
inspection.
(\ref{fig:leverage:zachary})
The so-called Zachary karate club network~\cite{zachary77karate}, with edges
color-coded such that leverage scores for a given edge increase from yellow
to red.
(\ref{fig:leverage:cumlev})
Cumulative leverage (with $k=10$) for a $65,031 \times 92,133$ term-document 
matrix constructed Enron electronic mail collection, illustrating 
that there are a large number of data points with very high leverage score.
(\ref{fig:leverage:bio})
The normalized statistical leverage scores and information gain 
score---information gain is a mutual information-based metric popular in 
the application area~\cite{Paschou07b,CUR_PNAS}---for each of the $n = 5520$ 
genes, a situation in which the data cluster well in the low-dimensional 
space defined by the maximum variance axes of the data~\cite{CUR_PNAS}.
Red stars indicate the $12$ genes with the highest leverage scores, and the 
red dashed line indicates the average or uniform leverage scores.
Note the strong correlation between the unsupervised leverage score metric 
and the supervised information gain metric.
}
\label{fig:leverage}
\end{figure}

To gain insight into these statistical leverage scores, consider the 
so-called ``wood beam data'' example~\cite{DS66,HW78}, which is visually 
presented in Figure~\ref{fig:leverage:wooddata}, along with the best-fit 
line to that data. 
In Figure~\ref{fig:leverage:woodscores}, the leverage scores for these ten 
data points are shown.
Intuitively, data points that ``stick out'' have particularly high 
leverage---\emph{e.g.}, the data point that has the most influence or 
leverage on the best-fit line to the wood beam data is the point marked 
``4'', and this is reflected in the relative magnitude of the corresponding 
statistical leverage score.
Indeed, since $\mbox{Trace}(H)=n$, a rule of thumb that has been suggested in
diagnostic regression analysis to identify errors and outliers in a data 
set is to investigate the $i^{th}$ data point if 
$H_{ii} > 2n/m$~\cite{VW81,ChatterjeeHadiPrice00}, 
\emph{i.e.}, if $H_{ii}$ is larger that $2$ or $3$ times the ``average'' 
size.
On the other hand, of course, if it happens to turn out that such a point is 
a legitimate data point, 
then one might expect that such an outlying data point will be a 
particularly important or informative data point.


Returning to the algorithmic perspective, consider the following random 
sampling algorithm for the LS approximation 
problem~\cite{DMM06,DMM08_CURtheory_JRNL}.
Given a very overconstrained least-squares problem, where the input matrix
$A$ and vector $b$ are \emph{arbitrary}, but $m \gg n$:
\begin{itemize}
\item
Compute normalized statistical leverage scores $\{p_i\}_{i=1}^{m}$, where 
$p_i = ||U_{A}^{(i)}||_2^2/k$, where $U$ is the $m \times n$ matrix 
consisting of the left singular vectors of $A$.%
\footnote{More generally, recall that if $U$ is \emph{any} orthogonal matrix spanning the 
column space of $A$, then $H=P_U=UU^T$ and thus $H_{ii}=||U^{(i)}||_2^2$, 
\emph{i.e.}, the statistical leverage scores equal the Euclidean norm of the 
rows of any such matrix $U$~\cite{CUR_PNAS}.
Clearly, the columns of $U$ are orthonormal, but the rows of $U$ in general 
are not---they can be uniform if, \emph{e.g.}, $U$ consists of columns from 
a truncated Hadamard matrix; or extremely nonuniform if, \emph{e.g.}, the 
columns of $U$ come from a truncated Identity matrix; or anything in 
between.}
\item
Randomly sample and rescale $r=O(n \log n /\epsilon^2 )$ constraints, 
\emph{i.e.}, rows of $A$ and the corresponding elements of $b$, using these
scores as an importance sampling distribution.
\item
Solve (using any appropriate LS solver as a black box) the induced 
subproblem to obtain a vector $\tilde{x}_{opt}$.
\end{itemize}
Since this algorithm samples constraints and not variables, the 
dimensionality of the vector $\tilde{x}_{opt}$ that solves the subproblem is 
the same as that of the vector $x_{opt}$ that solves the original problem.
This algorithm is described in more detail 
in~\cite{DMM06,DMM08_CURtheory_JRNL},
where it is shown that relative error bounds of the form 
\begin{eqnarray*}
||b-A\tilde{x}_{opt}||_2 
   &\leq& (1+\epsilon) ||b-Ax_{opt}||_2  \hspace{2mm} \mbox{ and } \\
||x_{opt} - \tilde{x}_{opt}||_2
   &\leq& O(\epsilon) ||x_{opt}||_2  
\end{eqnarray*}
hold.
Even more importantly, this algorithm highlights that the essential 
nonuniformity structure for the worst-case analysis of the LS (and, as we 
will see, the related CSSP) problem is defined by the statistical leverage 
scores!
That is, the same ``outlying'' data points that the diagnostic regression 
analyst tends to investigate are those points that are biased toward by the 
worst-case importance sampling probability distribution.

Clearly, for this LS algorithm, which holds for arbitrary input $A$ and $b$, 
$O(mn^2)$ time suffices to compute the sampling probabilities; in 
addition, it has recently been shown that one can obtain a nontrivial 
approximation to them in $o(mn^2)$ time~\cite{Malik10_TR}.
For many applications, such as those described in subsequent subsections, 
spending time on the order of computing an exact or approximate basis for 
the top-$k$ singular subspace is acceptable, in which case immediate 
generalizations of this algorithm are of interest.
In other cases, one can preprocess the LS system with a ``randomized 
Hadamard'' transform (as introduced in the ``fast'' Johnson-Lindenstrauss 
lemma~\cite{AC06,Matousek08_RSA}). 
Application of such a Hadamard transform tends to ``uniformize'' the 
leverage scores, intuitively for the same reason that a Fourier matrix 
delocalizes a localized $\delta$-function, in much the same way as 
application of a random orthogonal matrix or a random projection does.
This has led to the development of relative-error approximation algorithms 
for the LS problem that run in $o(mn^2)$ time in 
theory~\cite{Sarlos06,DMMS07_FastL2_TR}---essentially 
$O( mn \log(n/\epsilon) + \frac{n^3 \log^2 m}{\epsilon} )$ time, which is 
much less than $O(mn^2)$ when $m \gg n$---and whose numerical implementation
performs faster than traditional deterministic algorithms for systems with as
few as thousands of constraints and hundreds of 
variables~\cite{RT08,AMT09_DRAFT}.

\subsection{A two-stage hybrid algorithm for the CSSP}

In this subsection, I will describe an algorithm for the CSSP that uses the 
concept of statistical leverage to combine the NLA and TCS approaches, that 
comes with worst-case performance guarantees, and that performs well in 
practical data analysis applications.
I should note that, prior to this algorithm, it was not immediately clear 
how to combine these two approaches.
For example, if one looks at the details of the pivot rules in the 
deterministic NLA methods, it isn't clear that keeping more columns will 
help at all in terms of reconstruction error.
Similarly, since there is a version of the ``coupon collecting'' problem at 
the heart of the usual TCS analysis, keeping fewer than $\Omega(k \log k)$ 
will fail with respect to this worst-case analysis.
Moreover, the obvious hybrid algorithm of first randomly sampling 
$O(k \log k)$ columns and then using a deterministic QR procedure to select 
exactly $k$ of those columns does not seem to perform so well (either in 
theory or in practice).

Consider the following more sophisticated version of a two-stage hybrid 
algorithm.
Given an arbitrary $m \times n$ matrix $A$ and rank parameter $k$:
\begin{itemize}
\item
(Randomized phase)
Let $V_k^T$ be \emph{any} $k \times n$ orthogonal matrix spanning the 
top-$k$ right singular subspace of $A$.
Compute the importance sampling probabilities $\{p_i\}_{i=1}^{n}$, where 
\begin{equation}
p_i=\frac{1}{k}||{V_k^T}^{(i)}||_2^2  .
\label{eqn:sampling-probs}
\end{equation}
Randomly select and rescale $c = O(k \log k)$ columns of $V_k^T$ according 
to these probabilities.
\item
(Deterministic phase)
Let $\tilde{V}^T$ be the $k \times O(k \log k)$ non-orthogonal matrix 
consisting of the down-sampled and rescaled columns of $V_k^T$.
Run a deterministic QR algorithm on $\tilde{V}^T$ to select exactly $k$ 
columns of $\tilde{V}^T$.
Return the corresponding columns of $A$.
\end{itemize}
In particular, note that both the original choice of columns in the first 
phase, as well as the application of the QR algorithm in the second phase, 
involve the matrix $V_k^T$, \emph{i.e.}, the matrix defining the relevant
non-uniformity structure over the columns of $A$ in the 
($1+\epsilon$)-relative-error 
algorithm~\cite{DMM08_CURtheory_JRNL,CUR_PNAS}, rather than the matrix $A$ 
itself, as is more traditional.%
\footnote{Note that QR (as opposed to the SVD) is \emph{not} performed in 
the second phase to speed up the computation of a relatively cheap part of 
the algorithm, but instead it is performed since the goal of the algorithm 
is to return actual columns of the input matrix.}
A more detailed description of this algorithm may be found 
in~\cite{BMD08_CSSP_TR}, were it is shown that with extremely high 
probability the following spectral%
\footnote{Note that to establish the spectral norm 
bound,~\cite{BMD08_CSSP_TR} used slightly more complicated (but still 
depending only on information in $V_k^T$) importance sampling probabilities, 
but this may be an artifact of the analysis.}
and Frobenius norm bounds hold:
\begin{eqnarray*}
||A-P_CA||_2 &\le& O(k^{3/4} \log^{1/2}(k) n^{1/2}) ||A-P_{U_k}A||_2  \\
||A-P_CA||_F &\le& O(k \log^{1/2} k) ||A-P_{U_k}A||_F
\end{eqnarray*}
Interestingly, the analysis of this algorithm makes critical use of the 
importance sampling probabilities~(\ref{eqn:sampling-probs}), which are a 
generalization of the concept of \emph{statistical} leverage described in 
the previous subsection, for its worst-case \emph{algorithmic} performance 
guarantees.
Moreover, it is critical to the success of this algorithm that the QR 
procedure in the second phase be applied to the randomly-sampled version 
of $V_k^T$, \emph{i.e.}, the matrix defining the worst-case nonuniformity 
structure in $A$, rather than of $A$ itself.

With respect to running time, the computational bottleneck for this 
algorithm is computing $\{p_i\}_{i=1}^{n}$, for which it suffices to 
compute \emph{any} $k \times n$ matrix $V_k^T$ that spans the top-$k$ right 
singular subspace of $A$.
(In particular, a full SVD computation is \emph{not} necessary.)
Thus, this running time is of the same order as the running time of the QR 
algorithm used in the second phase when applied to the original matrix $A$.
Moreover, this algorithm easily scales up to matrices with thousands of rows
and millions of columns, whereas existing off-the-shelf implementations of 
the traditional algorithm may fail to run at all.
With respect to the worst-case quality of approximation bounds, this 
algorithm selects columns that are comparable to the state-of-the-art 
algorithms for constant $k$ (\emph{i.e.}, $O(k^{1/4}\log^{1/2}k)$ worse than 
previous work) for the spectral norm and only a factor of at most 
$O((k \log k)^{1/2})$ worse than the best previously-known existential 
result for the Frobenius norm.

\subsection{Data applications of the CSSP algorithm}
\label{sxn:cssp-data}

In the applications we have 
considered~\cite{Paschou07a,Paschou07b,CUR_PNAS,BMD08_CSSP_KDD,BMD09_kmeans_NIPS}, 
the goals of DNA microarray and DNA SNP analysis include the reconstruction 
of untyped genotypes, the evaluation of tagging SNP transferability between 
geographically-diverse populations, the classification and clustering into 
diseased and non-diseased states, and the analysis of admixed populations 
with non-trivial ancestry; and the goals of selecting good columns more 
generally include diagnostic data analysis and unsupervised feature 
selection for classification and clustering problems.
Here, I will give a flavor of when and why and how the CSSP algorithm of the 
previous subsection might be expected to perform well in these and other 
types of data applications.

To gain intuition for the behavior of leverage scores in a typical 
application, consider Figure~\ref{fig:leverage:zachary}, which illustrates
the so-called Zachary karate club network~\cite{zachary77karate}, a small 
but popular network in the community detection literature.
Given such a network $G=(V,E)$, with $n$ nodes, $m$ edges, and corresponding 
edge weights $w_e \ge 0$, define the $n \times n$ Laplacian matrix as 
$L=B^T W B$, where $B$ is the $m \times n$ edge-incidence matrix and $W$ is 
the $m \times m$ diagonal weight matrix.
The effective resistance between two vertices is given by the diagonal 
entries of the matrix $R= B L^{\dagger} B^T$ (where $L^{\dagger}$ denotes 
the Moore-Penrose generalized inverse) and is related to notions of 
``network betweenness''~\cite{newman05_betweenness}.
For many large graphs, this and related betweenness measures tend to be strongly correlated with
node degree and tend to be large for edges that form articulation points
between clusters and communities, \emph{i.e.}, for edges that ``stick out'' 
a lot.
It can be shown that the effective resistances of the edges of $G$ are 
proportional to the statistical leverage scores of the $m$ rows of the 
$m \times n$ matrix $W^{1/2}B$---consider the $m \times m$ matrix
$$ P = W^{1/2}RW^{1/2} =  \Phi(\Phi^T\Phi)^+\Phi^T ,$$
where $ \Phi = W^{1/2}B $, and note that if $U_{\Phi}$ denotes any 
orthogonal matrix spanning the column space of $\Phi$, then
$$ P_{ii} = (U_{\Phi}U_{\Phi}^T)_{ii} = ||(U_{\Phi})_{(i)}||_2^2 .$$
Figure~\ref{fig:leverage:zachary} presents a color-coded illustration of 
these scores for Zachary karate club network.

Next, to gain intuition for the (non-)uniformity properties of statistical 
leverage scores in a typical application, consider a term-document matrix 
derived from the publicly-released 
Enron electronic mail collection~\cite{BB06}, which is an example of social 
or information network we will encounter again in the next section and which 
is also typical of the type of data set to which SVD-based latent semantic 
analysis (LSA) methods~\cite{DDLFH90} have been applied.
I constructed a $65,031 \times 92,133$ matrix, as described in~\cite{BB06}, 
and I chose the rank parameter as $k=10$.
Figure~\ref{fig:leverage:cumlev} plots the cumulative leverage, \emph{i.e.}, 
the running sum of top $t$ statistical leverage scores, as a function of 
increasing $t$.
Since $\frac{k}{n}=\frac{10}{92,133}\approx1.0854\times10^{-4}$, we see that 
the highest leverage term has a leverage score nearly two orders of 
magnitude larger than this ``average'' size scale, that the second 
highest-leverage score is only marginally less than the first, that the 
third highest score is marginally less than the second, etc.
Thus, by the traditional metrics of diagnostic data 
analysis~\cite{VW81,ChatterjeeHadiPrice00}, which suggests flagging a data 
point if 
$$ (P_{U_k})_{ii} = (H_k)_{ii}> 2k/n ,$$
there are a \emph{huge} number of data points that are \emph{extremely} outlying.
In retrospect, of course, this might not be surprising since the Enron email 
corpus is extremely sparse, with nowhere on the order of $\Omega(n)$ 
nonzeros per row.
Thus, even though LSA methods have been successfully applied, plausible 
generative models associated with these data are clearly not Gaussian, and 
the sparsity structure is such that there is no reason to expect that nice 
phenomena such as measure concentration occur.

Finally, note that DNA microarray and DNA SNP data often exhibit a similar 
degree of nonuniformity, although for somewhat different reasons.
To illustrate, Figure~\ref{fig:leverage:bio} presents two plots.
First, it plots the normalized statistical leverage scores 
for a data matrix, as was described in~\cite{CUR_PNAS}, consisting of 
$m = 31$ patients with $3$ different cancer types with respect to $n = 5520$ 
genes. 
A similar plot illustrating the remarkable nonuniformity in statistical 
leverage scores for DNA SNP data was presented in~\cite{Paschou07b}.
Empirical evidence suggests that two phenomena may be responsible.
First, as with the term-document data, there is no domain-specific reason to 
believe that nice properties like measure concentration occur---on the 
contrary, there are reasons to expect that they do not.
Recall that each DNA SNP corresponds to a single mutational event in human 
history.
Thus, it will ``stick out,'' as its description along its one axis in the 
vector space will likely not be well-expressed in terms of the other axes, 
\emph{i.e.}, in terms of the other SNPs, and by the time it ``works its way 
back'' due to population admixing, etc., other SNPs will have occurred 
elsewhere.
Second, the correlation between statistical leverage and supervised mutual 
information-based metrics is particularly prominent in examples where the 
data cluster well in the low-dimensional space defined by the maximum 
variance axes.
Considering such data sets is, of course, a strong selection bias, but it is 
common in applications.
It would be of interest to develop a model that quantifies the observation
that, conditioned on clustering well in the low-dimensional space, an 
unsupervised measure like leverage scores should be expected to correlate 
well with a supervised measure like informativeness~\cite{Paschou07b} or
information gain~\cite{CUR_PNAS}.

With respect to some of the more technical and implementational issues,
several observations~\cite{BMD08_CSSP_KDD,BMD08_CSSP_TR} shed light on the 
inner workings of the CSSP algorithm and its usefulness in applications.
Recall that an important aspect of QR algorithms is how they make so-called
pivot rule decisions about which columns to keep~\cite{GVL96} and that such 
decisions can be tricky when the columns are not orthogonal or spread out in 
similarly nice ways.
\begin{itemize}
\item
We looked at several versions of the QR algorithm, and we compared each 
version of QR to the CSSP using that version of QR in the second phase.
One observation we made was that different QR algorithms behave 
differently---\emph{e.g.}, some versions such as the Low-RRQR algorithm 
of~\cite{CH94} tend to perform much better than other versions such as the 
qrxp algorithm of~\cite{BQ98a,BQ98b}.
Although not surprising to NLA practitioners, this observation indicates 
that some care should be paid to using ``off the shelf'' implementations in 
large-scale applications.
A second less-obvious observation is that preprocessing with the randomized 
first phase tends to improve more poorly-performing variants of QR more than 
better variants.
Part of this is simply that the more poorly-performing variants have more 
room to improve, but part of this is also that more sophisticated versions 
of QR tend to make more sophisticated pivot rule decisions, which are 
relatively less important after the randomized bias toward directions that 
are ``spread out.''
\item
We also looked at selecting columns by applying QR on $V_k^T$ and then 
keeping the corresponding columns of $A$, \emph{i.e.}, just running the 
classical deterministic QR algorithm with no randomized first phase on the 
matrix $V_k^T$.
Interestingly, with this ``preprocessing'' we tended to get better columns 
than if we ran QR on the original matrix $A$.
Again, the interpretation seems to be that, since the norms of the columns of 
$V_k^T$ define the relevant nonuniformity structure with which to sample with
respect to, working directly with those columns tends make things ``spread 
out,'' thereby avoiding (even in traditional deterministic settings) 
situations where pivot rules have problems.
\item
Of course, we also observed that randomization further improves the results, 
assuming that care is taken in choosing the rank parameter $k$ and the 
sampling parameter $c$.
In practice, the choice of $k$ should be viewed as a ``model selection'' 
question.
Then, by choosing $c=k,1.5k,2k,\ldots$, we often observed a 
``sweet spot,'' in bias-variance sense, as a function of increasing $c$.
That is, for a fixed $k$, the behavior of the deterministic QR algorithms 
improves by choosing somewhat more than $k$ columns, but that improvement 
is degraded by choosing too many columns in the randomized phase.
\end{itemize}

\subsection{Some general thoughts on leverage scores and matrix algorithms}

I will conclude this section with two general observations raised by these
theoretical and empirical results having to do with using the concept of 
statistical leverage to obtain columns from an input data matrix that are
good both in worst-case analysis and also in large-scale data applications.

One high-level question raised by these results is: why should statistical 
leverage, a traditional concept from regression diagnostics, be useful to 
obtain improved worst-case approximation algorithms for traditional NLA 
matrix problems?
The answer to this seems to be that, intuitively, if a data point has a high 
leverage score and is not an error then it might be a particularly 
important or informative data point.  
Since worst-case analysis takes the input matrix as given, each row is 
assumed to be reliable, and so worst-case guarantees are obtained by 
focusing effort on the most informative data points.
It would be interesting to see if this perspective is applicable more 
generally in the design of matrix and graph algorithms.

A second high-level question is: why are the statistical leverage scores 
so nonuniform in many large-scale data analysis applications.
Here, the answer seems to be that, intuitively, in many very large-scale 
applications, statistical models are \emph{implicitly} assumed based on 
computational and not statistical considerations.
In these cases, it is not surprising that some interesting data points 
``stick out'' relative to obviously inappropriate models.
This suggests the use of these importance sampling scores as cheap 
signatures of the ``inappropriateness'' of a statistical model (chosen for 
algorithmic and not statistical reasons) in large-scale exploratory or
diagnostic applications.
It would also be interesting to see if this perspective is applicable more 
generally.

\section{Internet applications and novel graph algorithms}

In this section, I will describe a novel perspective on identifying good
clusters or communities in a large graph.
The general problem of finding good clusters in (or good partitions of) a data graph
has been studied extensively in a wide range of applications. 
For example, it has been studied for years in scientific computation (where 
one is interested in load balancing in parallel computing applications),
machine learning and computer vision (where one is interested in segmenting 
images and clustering data), and theoretical computer science (where one is 
interested in it as a primitive in divide-and-conquer algorithms).
More recently, problems of this sort have arisen in the analysis of large 
social and information networks, where one is interested in finding 
communities that are meaningful in a domain-specific context.

\subsection{Motivating Internet application}

Sponsored search is a type of contextual advertising where Web site owners 
pay a fee, usually based on click-throughs or ad views, to have their Web 
site search results shown in top placement position on search engine result 
pages.
For example, when a user enters a term into a search box, the search engine
typically presents not only so-called ``algorithmic results,'' but it also 
presents text-based advertisements that are important for revenue 
generation.
In this context, one can construct a so-called \emph{advertiser-bidded-phrase
graph}, the simplest variant of which is a bipartite graph $G=(U,V,E)$, in 
which $U$ consists of some discretization of the set of advertisers, $V$ 
consists of some discretization of the set of keywords that have been bid
upon, and an edge $e=(u,v) \in E$ is present if advertiser $u \in U$ had bid 
on phrase $v \in V$.
It is then of interest to perform data mining on this graph in order to 
optimize quantities such as the user click-through-rate or advertiser 
return-on-investment.

Numerous community-related problems arise in this context.
For example, in \emph{micro-market identification}, one is interested in 
identifying a set of nodes that is large enough that it is worth spending an
analyst's time upon, as well as coherent enough that it can be usefully 
thought about as a ``market'' in an economic sense.
Such a cluster can be useful for A/B bucket testing, as well as for 
recommending to advertisers new queries or sub-markets.
Similarly, in \emph{advanced match}, an advertiser places bids not only when 
an exact match occurs to a set of advertiser-specified phrases, but also
when a match occurs to phrases ``similar to'' the specified bid phrases.
Ignoring numerous natural language and game theoretic issues, one can 
imagine that if the original phrase is in the middle of a fairly homogeneous 
concept class, then there may be a large number of similar phrases that are 
nearby in the graph topology, in which case it might be advantageous to both 
the advertiser and the search engine to include those phrases in the 
bidding.
On the other hand, if the original phrase is located in a locally very 
unstructured part of the graph, then there may be a large number of phrases 
that are nearby in the graph topology but that have a wide range of meanings 
very different than the original bid phrase, in which case performing such 
an expansion might not make sense.

\begin{figure}
   \begin{center}
   \begin{tabular}{ccc}
      \subfigure[A toy network]{
         \includegraphics[width=0.20\textwidth]{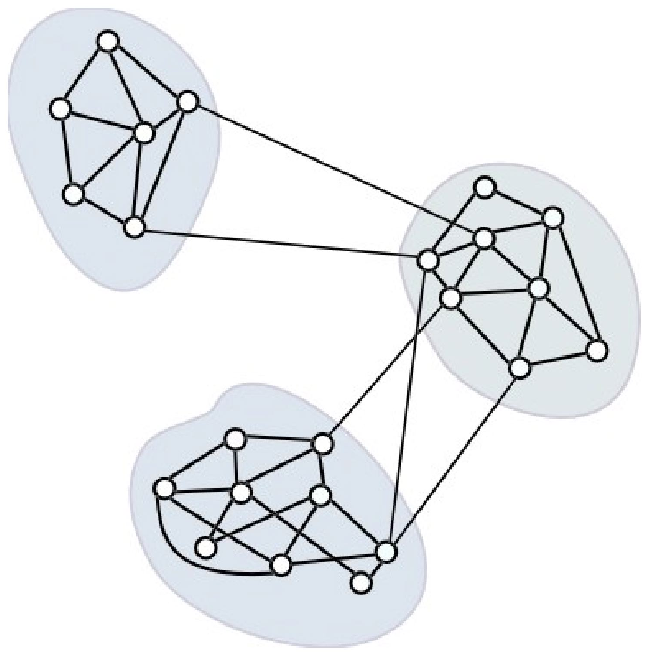}
         \label{fig:communities:toy-kmeans}
      } &
      \subfigure[A schematic illustration]{
         \includegraphics[width=0.35\textwidth]{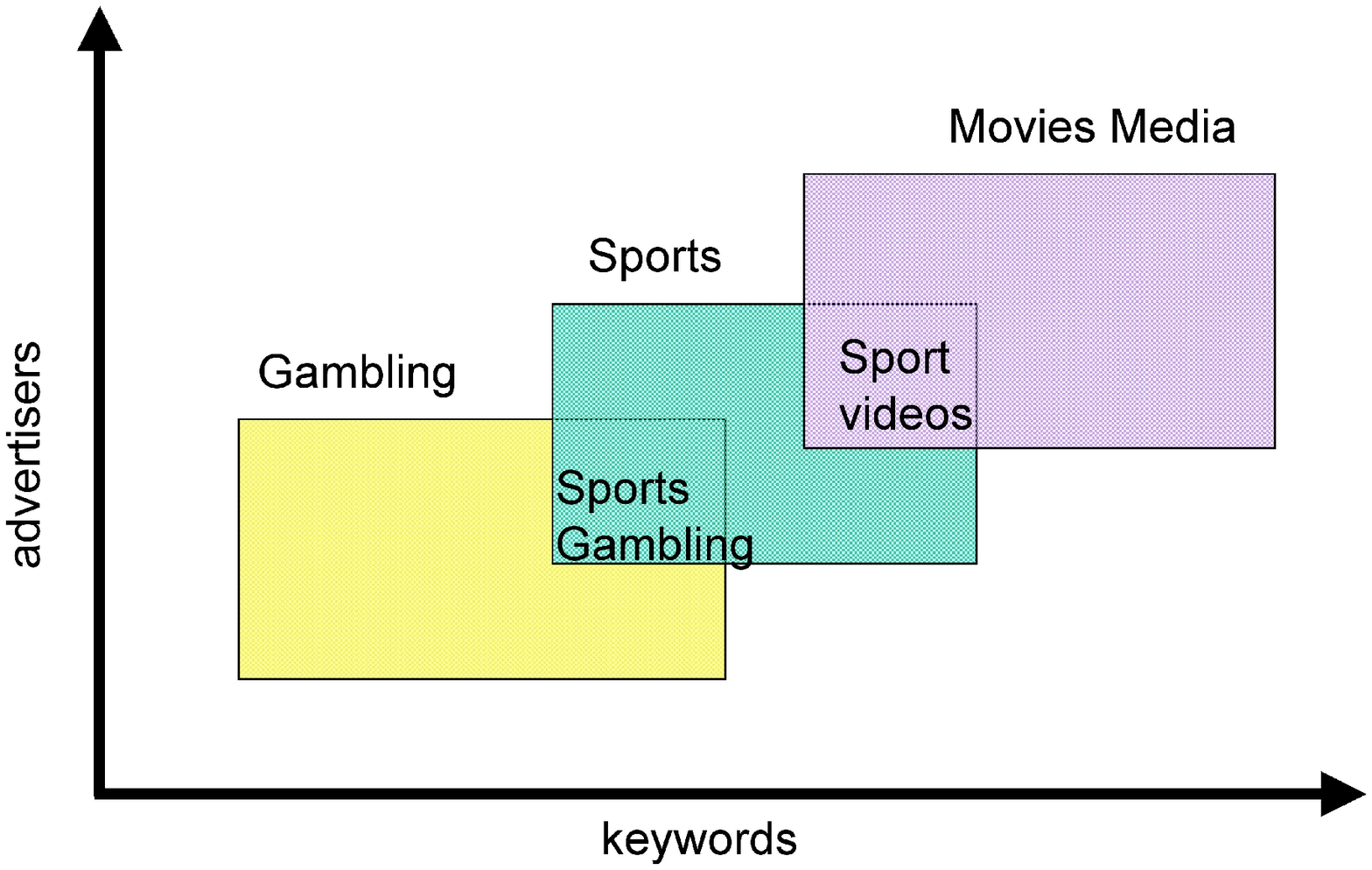}
         \label{fig:communities:ad-bid-schematic}
      } &
      \subfigure[A realistic network]{
         \includegraphics[width=0.35\textwidth]{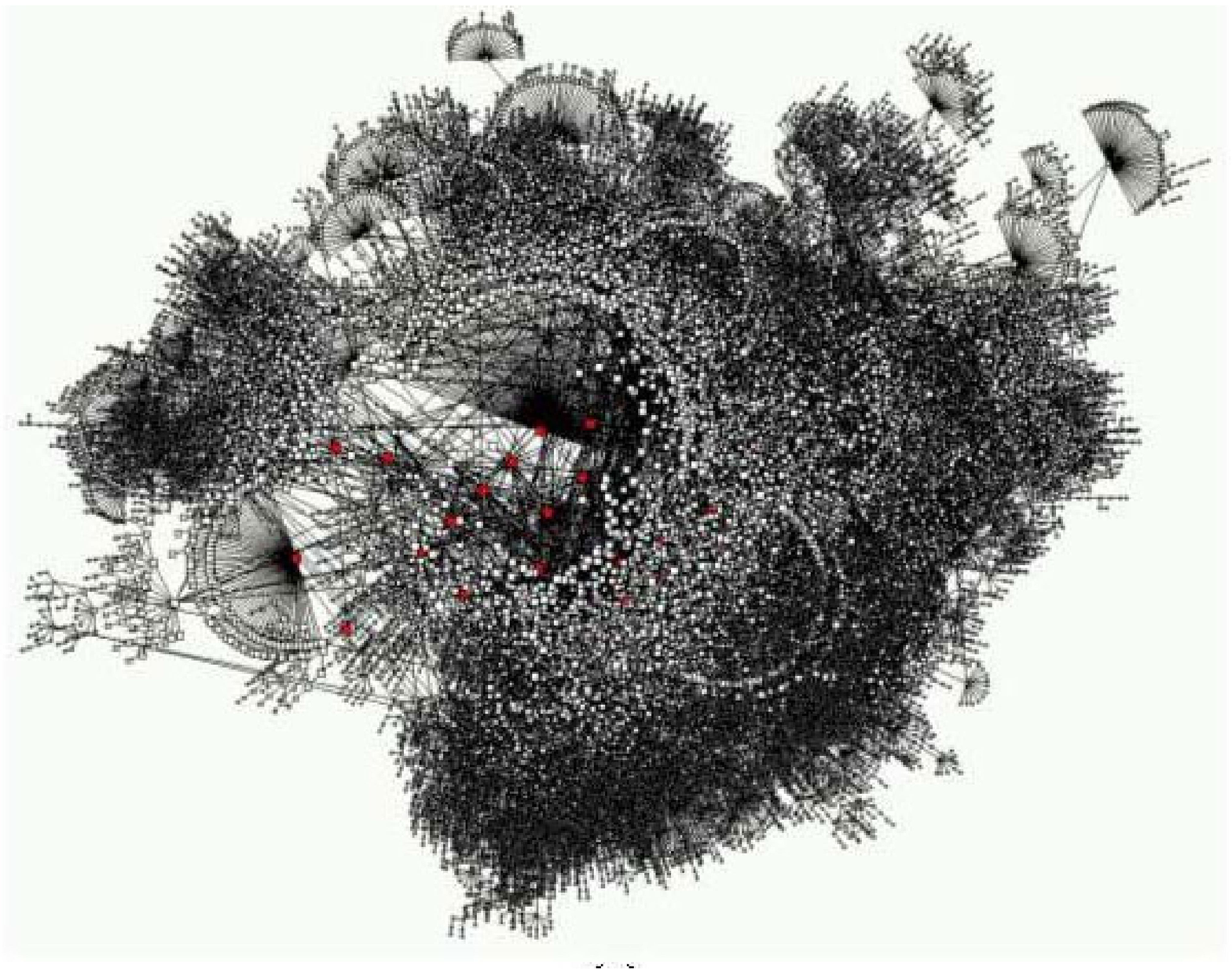}
         \label{fig:communities:real-network-vis}
      } \\
      \subfigure[Clusters of different sizes in a small network]{
         \includegraphics[width=0.20\textwidth]{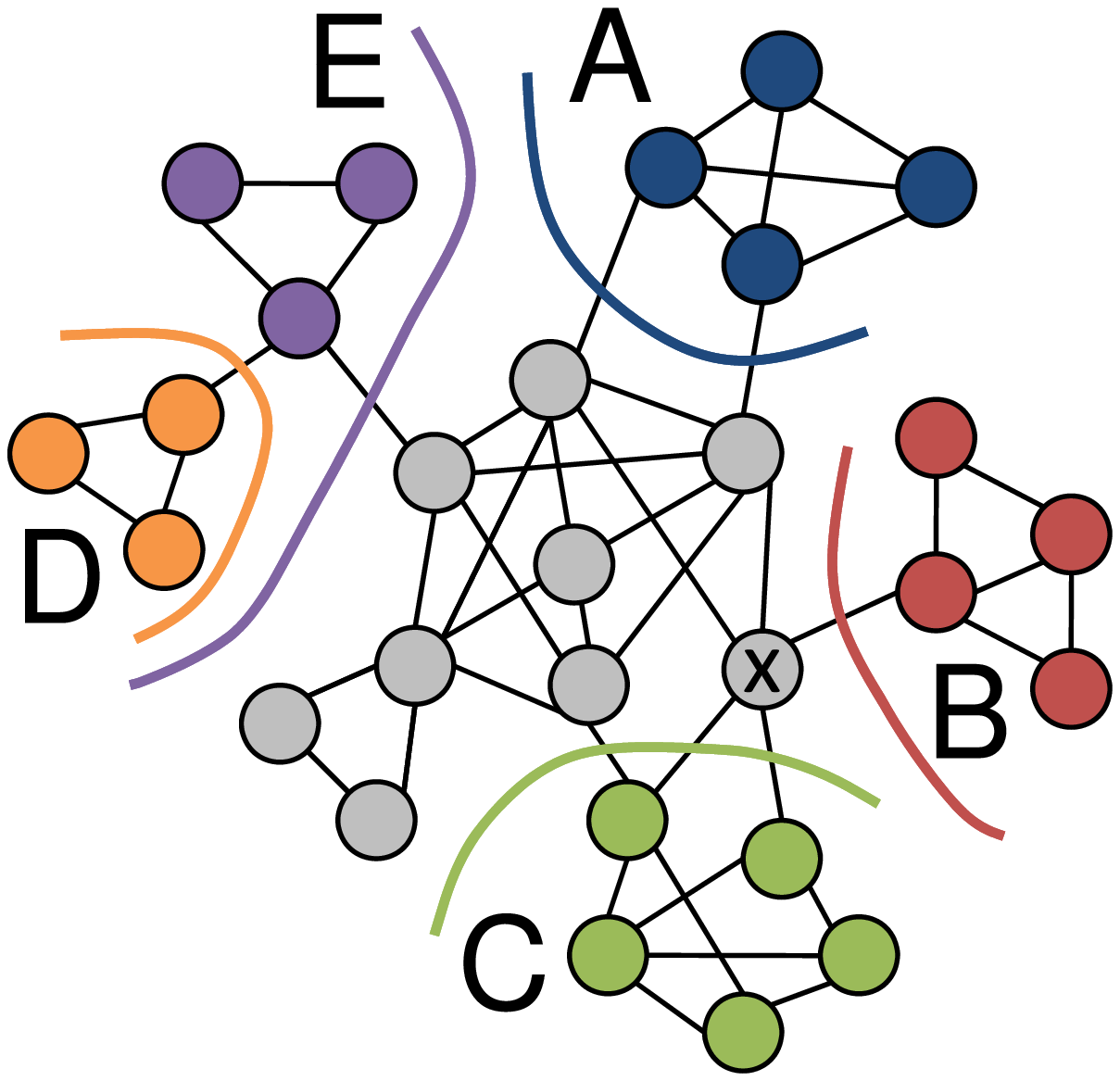}
         \label{fig:communities:toy-fig2}
      } &
      \subfigure[NCP of the small network]{
         \includegraphics[width=0.30\textwidth]{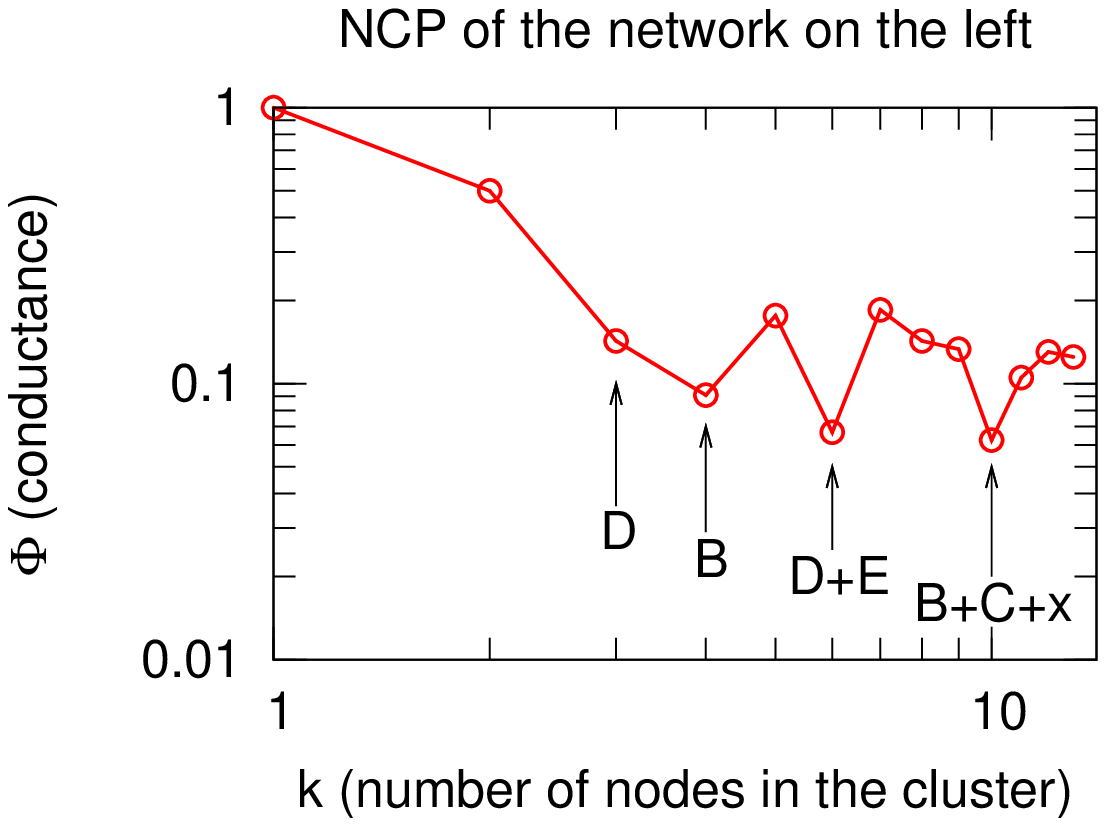}
         \label{fig:communities:toy-fig2-ncp}
      } &
      \subfigure[NCP of a realistic network]{
         \includegraphics[width=0.30\textwidth]{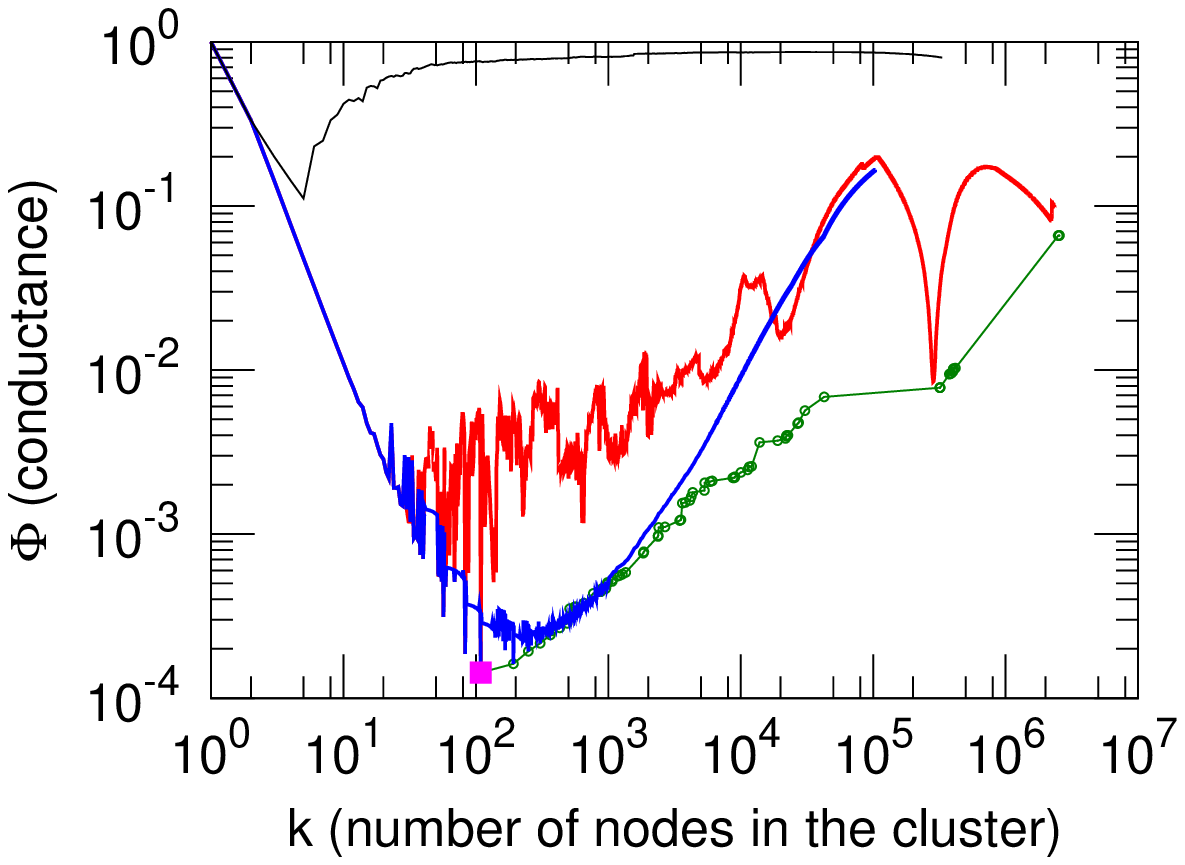}
         \label{fig:communities:realNCP-LiveJournal}
      } \\
      \subfigure[NCP of another realistic network]{
         \includegraphics[width=0.30\textwidth]{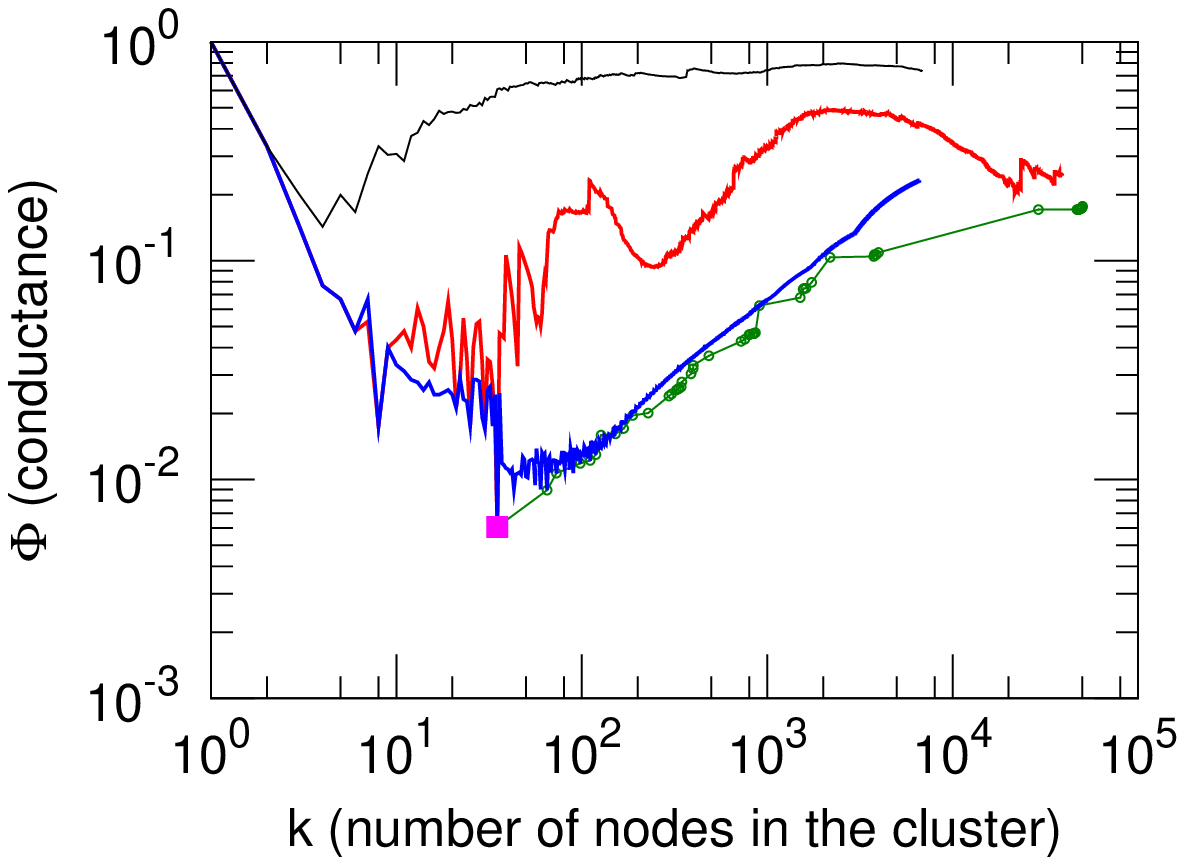}
         \label{fig:communities:realNCP-epinions}
      } &
      \subfigure[Nested core-periphery structure]{
         \includegraphics[width=0.35\textwidth]{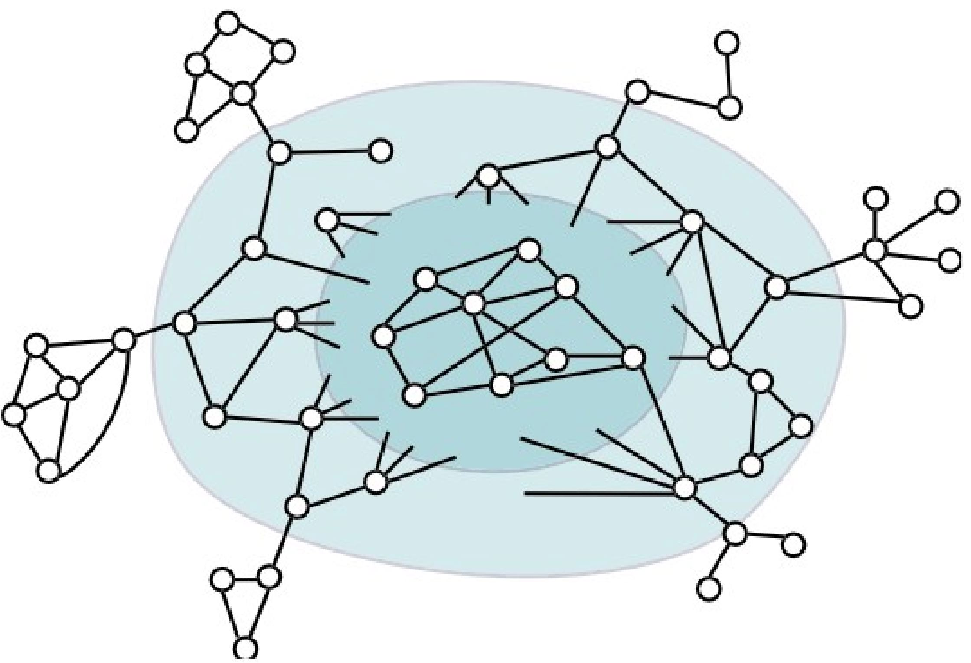}
         \label{fig:communities:core-periphery}
      } &
      \subfigure[A ``spectral'' community]{
         \includegraphics[width=0.25\textwidth]{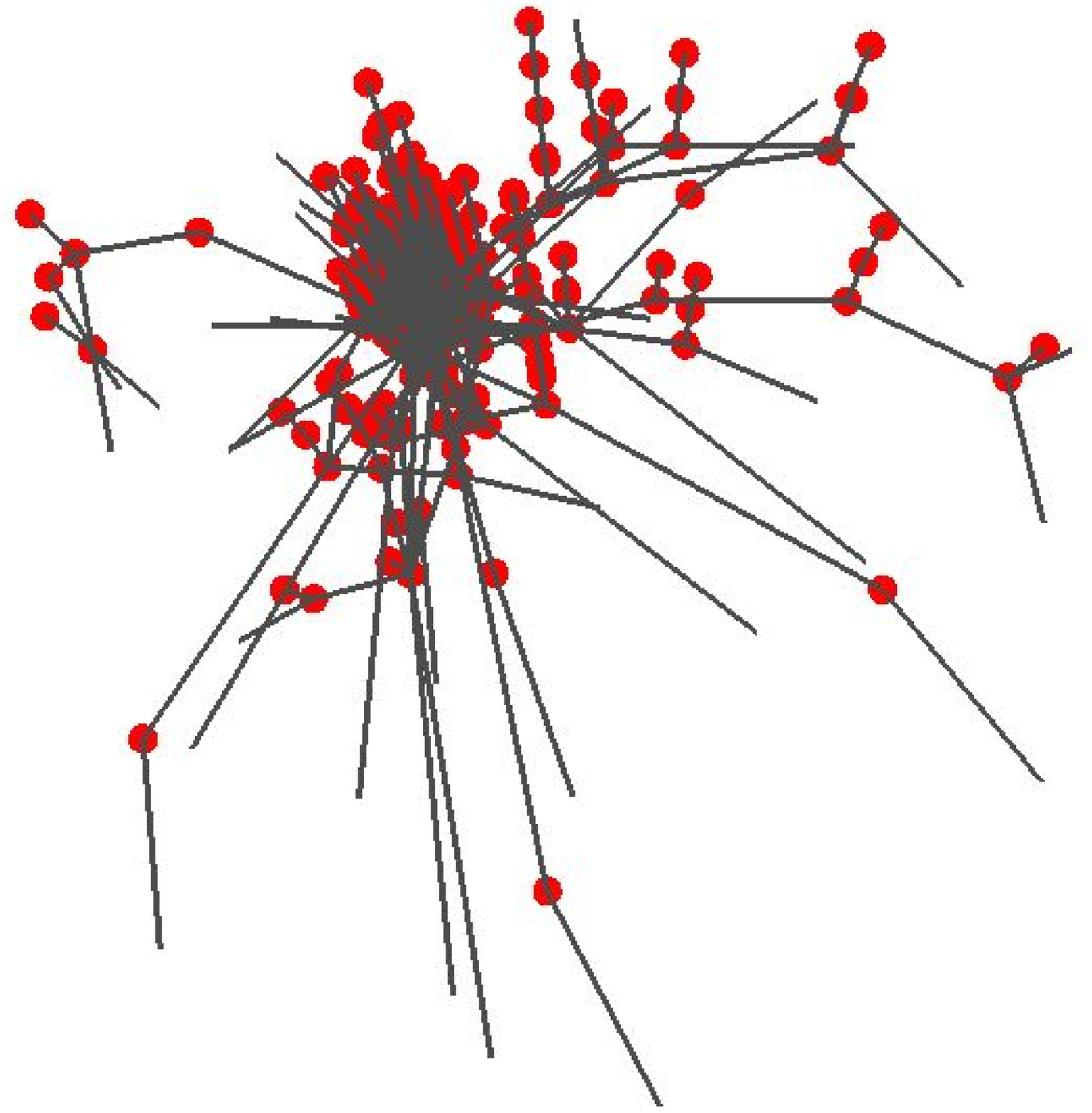}
         \label{fig:communities:spectral-cmty-1}
      } \\
      \subfigure[Another ``spectral'' community]{
         \includegraphics[width=0.25\textwidth]{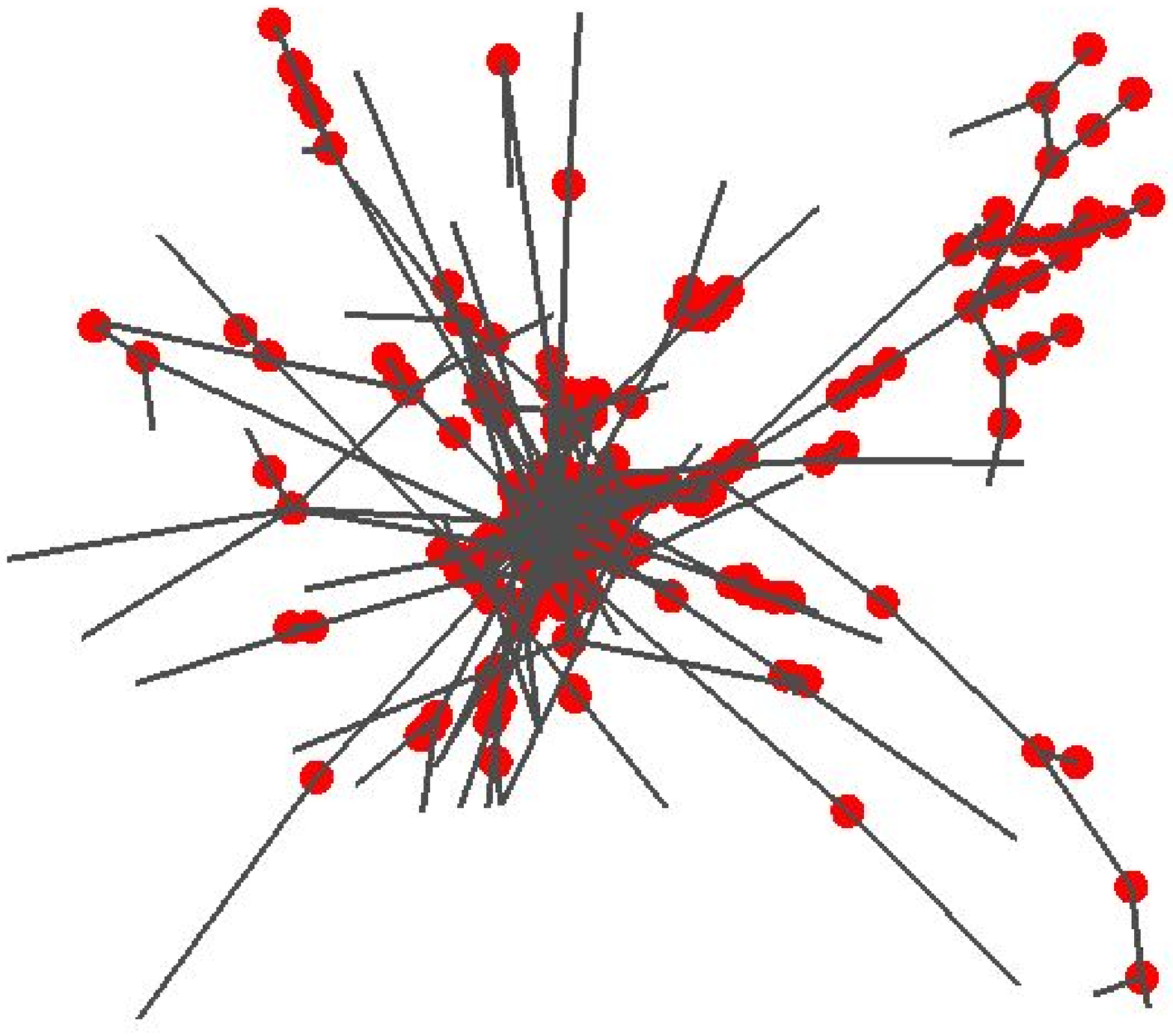}
         \label{fig:communities:spectral-cmty-2}
      } &
      \subfigure[A ``flow'' community]{
         \includegraphics[width=0.25\textwidth]{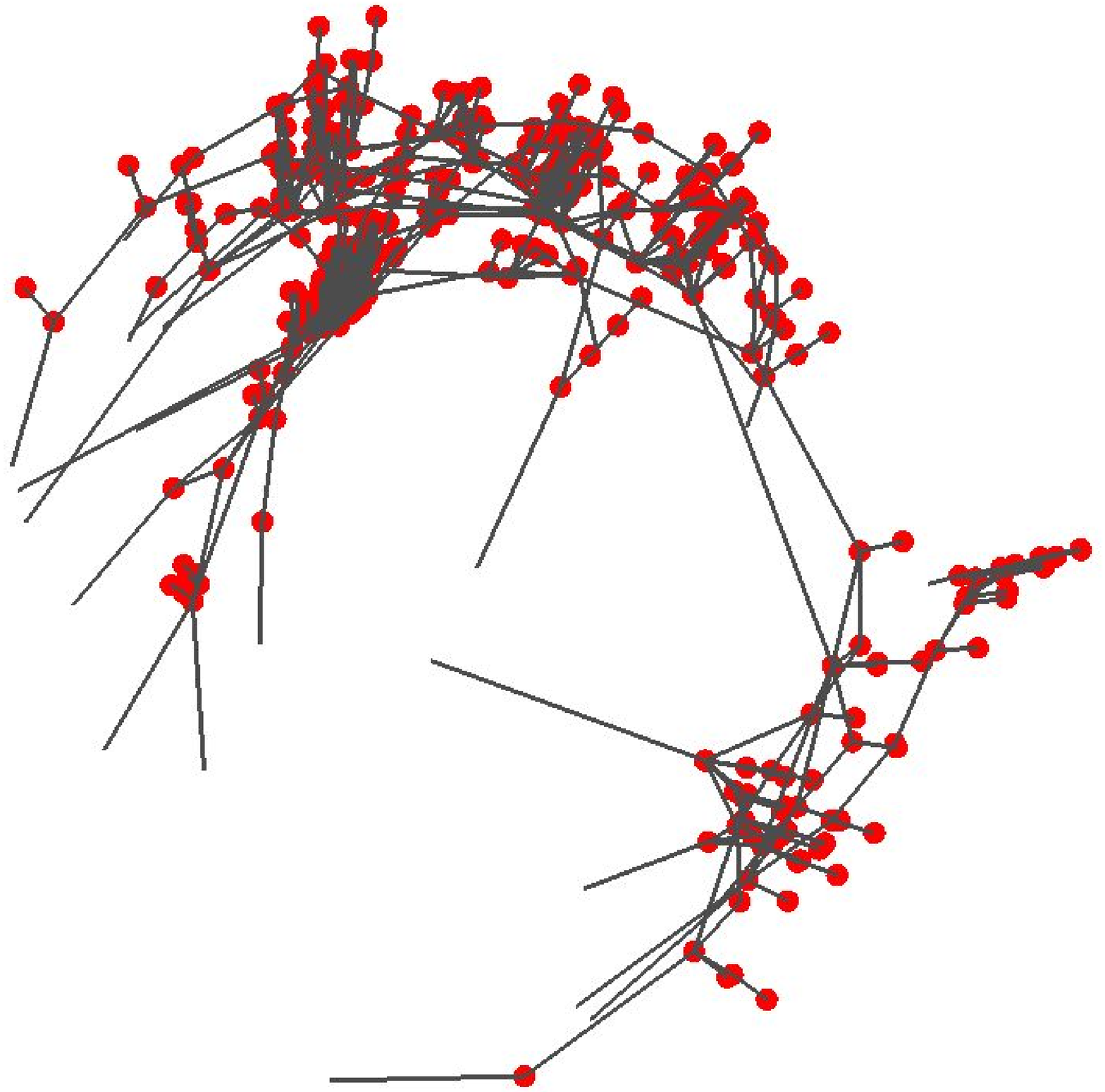}
         \label{fig:communities:flow-cmty-1}
      } &
      \subfigure[Another ``flow'' community]{
         \includegraphics[width=0.25\textwidth]{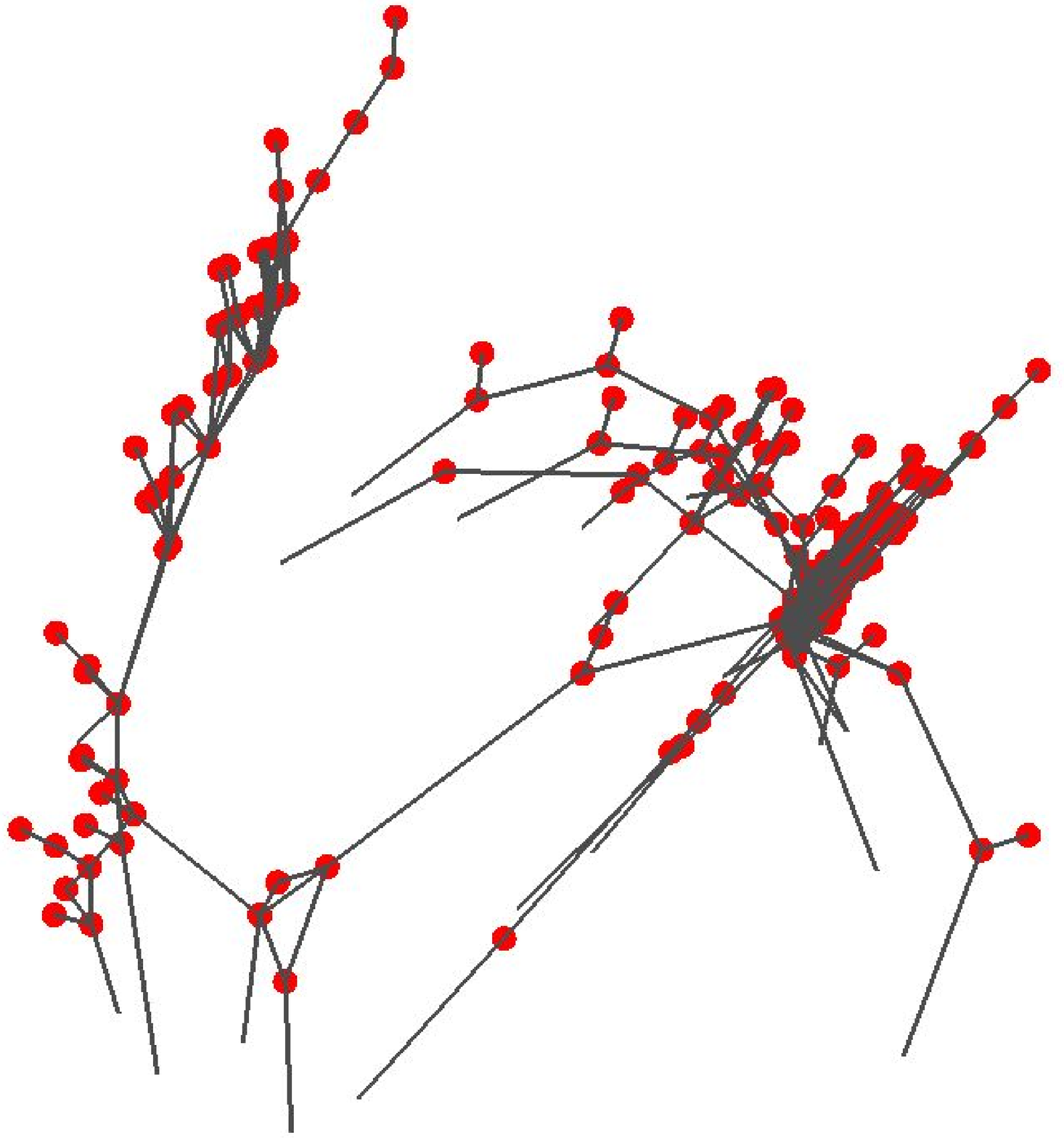}
         \label{fig:communities:flow-cmty-2}
      } 
   \end{tabular}
   \end{center}
\caption{
(\ref{fig:communities:toy-kmeans})
A toy network that is ``nice'' and easily-visualizable.
(\ref{fig:communities:ad-bid-schematic})
A schematic of an advertiser-bidded-phrase graph that is similarly ``nice.''
(\ref{fig:communities:real-network-vis})
A more realistic depiction of a large network.
(\ref{fig:communities:toy-fig2})
Illustration of conductance clusters in a small network.
(\ref{fig:communities:toy-fig2-ncp})
The downward-sloping NCP common to small networks, low-dimensional graphs, 
and common hierarchical models.
(\ref{fig:communities:realNCP-LiveJournal})
The NCP of a ``LiveJournal'' network~\cite{LLDM08_communities_TR}.
(\ref{fig:communities:realNCP-epinions})
The NCP of an ``Epinions'' network~\cite{LLDM08_communities_TR}.
(\ref{fig:communities:core-periphery})
The nested core-periphery structure responsible for the upward-sloping NCP.
(\ref{fig:communities:spectral-cmty-1})
A typical $500$-node ``community'' found with a spectral method.
(\ref{fig:communities:spectral-cmty-2})
Another $500$-node ``community'' found with a spectral method.
(\ref{fig:communities:flow-cmty-1})
A typical $500$-node ``community'' found with a flow-based method.
(\ref{fig:communities:flow-cmty-2})
Another $500$-node ``community'' found with a flow-based method.
}
\label{fig:communities}
\end{figure}

As in many other application areas, 
in these clustering and community identification applications, 
a common \emph{modus operandi} in applying data analysis tools 
is:
\begin{itemize}
\item
Define an objective function that formalizes the intuition that one has as 
a data analyst as to what constitutes a good cluster or community.
\item
Since that objective function is almost invariably intractable to optimize 
exactly, apply some approximation algorithm or heuristic to the problem.
\item
If the set of nodes that are thereby obtained 
look plausibly good in an application-dependent sense, then declare success.
Otherwise, modify the objective function or heuristic or algorithm, and 
iterate the process.
\end{itemize}
Such an approach can lead to insight---\emph{e.g.}, if the data are 
``morally'' low-dimensional, as might be the case, as illustrated in 
Figure~\ref{fig:communities:toy-kmeans}, when the Singular Value 
Decomposition, manifold-based machine learning methods, or $k$-means-like 
statistical modeling assumptions are appropriate; 
or if the data have other ``nice'' hierarchical properties that conform to 
one's intuition, as suggested by the schematic illustration in 
Figure~\ref{fig:communities:ad-bid-schematic};
or if the data come from a nice generative model such as a mixture model.
In these cases, a few steps of such a procedure will likely lead to a 
reasonable solution.

On the other hand, if the size of the data is larger, or if the data arise 
from an application where the sparsity structure and noise properties are 
more adversarial and less intuitive, then such an approach can be 
problematic.
For example, if a reasonable solution is not readily obtained, then it is 
typically not clear whether one's original intuition was wrong; whether this 
is due to an improper formalization of the correct intuitive concept; 
whether insufficient computational resources were devoted to the problem;
whether the sparsity and noise structure of the data have been modeled 
correctly, etc.
That common visualization algorithms applied to large networks lead to 
largely non-interpretable figures, as illustrated in 
Figure~\ref{fig:communities:real-network-vis}, which reveal more about the 
inner workings of the visualization algorithm than the network being 
visualized, \emph{suggests} that this more problematic situation is more 
typical of large social and information network data.
In such cases, it would be of interest to have principled tools, the 
algorithmic and statistical properties of which are well-understood, to 
``explore'' the data.

\subsection{A formalization of and prior approaches to this problem}

One of the most basic question that one can ask about a data set (and one 
which is intimately related to questions of community identification) is: 
what does the data set ``look like'' if it is cut into two pieces?
This is of interest since, \emph{e.g.}, one would expect very different 
properties from a data set that ``looked like'' a hot dog or a pancake, in 
that one could split it into two roughly equally-sized pieces, each of which 
was meaningfully coherent in and of itself; 
as opposed to a data set that ``looked like'' a moderately-dense 
expander-like random graph, in that there didn't exist any good partitions 
of any size of the data into two pieces; as opposed to a data set in which 
there existed good partitions, but those involved nibbling off just $0.1\%$ 
of the nodes of the data and leaving the rest intact.

A common way to formalize this question of qualitative connectivity is via 
the \emph{graph partitioning}
problem~\cite{Pot96,jain99data,ShiMalik00_NCut,gaertler05_clustering,luxburg05_survey,newman2006finding,Schaeffer07_survey}.
Graph partitioning refers to a family of objective functions and associated
approximation algorithms that involve cutting or partitioning the nodes of a 
graph into two sets with the goal that the cut has good quality 
(\emph{i.e.}, not much edge weight crosses the cut) as well as good balance 
(\emph{i.e.}, each of the two sets has a lot of the node weight).
There are several standard formalizations of this bi-criterion. 
In this chapter, I will be interested in the quotient cut formulations,%
\footnote{Other formulations include the graph bisection problem, which 
requires perfect $\frac{1}{2}:\frac{1}{2}$ balance and the $\beta$-balanced 
cut problem, with $\beta$ set to a fraction such as $\frac{1}{10}$, which 
requires at least a $\beta:(1-\beta)$ balance.}
which require the small side to be large enough to ``pay for'' the edges 
in the cut.

Given an undirected, possibly weighted, graph $G=(V,E)$, the 
\textit{expansion $\alpha(S)$ of a set of nodes $S \subseteq V$} is:
\begin{equation}
\alpha(S) = \frac{|E(S, \overline{S})|}{\min\{|S|,|\overline{S}|)\}},
\label{eqn:expansion_set}
\end{equation}
where $E(S, \overline{S})$ denotes the set of edges having one end in $S$ 
and one end in the complement $\overline{S}$, and where $|\cdot|$ denotes 
cardinality (or weight); and the \textit{expansion of the graph $G$} is:
\begin{equation}
\alpha(G) = \min_{S \subseteq V} \alpha(S)  .
\label{eqn:expansion_graph}
\end{equation}
Alternatively, if there is substantial variability in node degree, then 
normalizing by a related quantity is of greater interest.
The \emph{conductance $\phi(S)$ of a set of nodes $S \subset V$} is: 
\begin{equation}
\phi(S) = \frac{ |E(S, \overline{S})| }{ \min\{A(S),A(\bar{S})\} } ,
\label{eqn:conductance_set}
\end{equation}
where $A(S) = \sum_{i \in S} \sum_{j \in V} A_{ij} $, where $A$ is the 
adjacency matrix of a graph.
In this case, the \textit{conductance of the graph $G$} is:
\begin{equation}
\phi(G) = \min_{S \subseteq V} \phi(S)  .
\label{eqn:conductance_graph}
\end{equation}
In either case, one could replace the ``min'' in the denominator with a 
``product,'' \emph{e.g.}, replace $\min\{A(S),A(\bar{S})\}$ in the 
denominator of Eqn.~(\ref{eqn:conductance_set}) with 
$A(S)\cdot A(\bar{S})$.%
\footnote{Note that, up to scaling, this is simply the popular Normalized 
Cute metric~\cite{ShiMalik00_NCut}.}
The product formulation provides a slightly greater reward to a cut for 
having a big big-side, and it so has a slightly weaker preference for 
balance than the minimum formulation.
Both formulations are equivalent, though, in that the objective function 
value of the set of nodes achieving the minimum with the one is within a 
factor of $2$ of the objective function value of the (in general different) 
set of nodes achieving the minimum with the other.
Generically, the Minimum Conductance Cut Problem refers to solving any of 
these formulations; and
importantly, all of these variants of the graph partitioning problem lead to 
intractable combinatorial optimization problems. 

Within scientific 
computing~\cite{Pot96,spielman96_spectral,karypis98_metis,karypis98metis}, 
and more recently within statistics and machine 
learning~\cite{ShiMalik00_NCut,weiss99_segmentation,NJW01_spectral,KVV04_JRNL},
a great deal of work has focused on the conductance (or normalized cut) 
versions of this graph partitioning problem.
Several general observations about the approach adopted in these literatures
include:
\begin{itemize}
\item
The focus is on \emph{spectral approximation algorithms}.
Spectral algorithms use an exact or approximate eigenvector of the graph's 
Laplacian matrix to find a cut that achieves a ``quadratic approximation,'' 
in the sense that the cut returned by the algorithm has conductance value no 
bigger than $\phi$ if the graph actually contains a cut with conductance 
$O(\phi^2)$~\cite{Cheeger69_bound,Donath:1972,fiedler73graphs,mohar91_survey,Chung:1997}.
\item
The focus is on \emph{low-dimensional graphs}, \emph{e.g.}, bounded-degree 
planar graphs and finite element meshes (for which 
quality-of-approximation bounds are known that depend on just the number 
of nodes and not on structural parameters such as the conductance value).
Moreover, there is an acknowledgement that \emph{spectral methods are 
inappropriate for expander graphs} that have constant expansion or 
conductance (basically, since for these graphs any clusters returned are not 
meaningful in applications).
\item
Since the algorithm is typically applied to find partitions of a graph that 
are useful in some downstream application, there is a \emph{strong interest 
in the actual pieces returned by the algorithm}.
Relatedly, there is interest in the robustness or consistency properties of 
spectral approximation algorithms under assumptions on the data.
\end{itemize}
Recursive bisection heuristics---recursively divide the graph into two 
groups, and then further subdivide the new groups until the desired number 
of clusters groups is achieved---are also common here.
They may be combined with local improvement 
methods~\cite{Kernighan:1970,Fiduccia:1982}, which when combined with 
multi-resolution ideas leads to programs such as 
Metis~\cite{karypis98_metis,karypis98metis}, Cluto~\cite{zhao04cluto}, and
Graclus~\cite{dhillon07graclus}.

Within TCS, a great deal of work has also focused on this graph partitioning
problem~\cite{Leighton:1988,LLR95_JRNL,Shm96,Leighton:1999,Arora:2004,ARV_CACM08}.
Several general observations about the traditional TCS approach include:
\begin{itemize}
\item
The focus is on \emph{flow-based approximation algorithms}.
These algorithms use multi-commodity flow and metric embedding ideas to 
find a cut whose conductance is within an $O(\log n)$ factor of 
optimal~\cite{Leighton:1988,LLR95_JRNL,Leighton:1999}, in the sense that the 
cut returned by the algorithm has conductance no bigger than $O(\log n)$, 
where $n$ is the number of nodes in the graph, times the conductance value 
of the optimal conductance set in the graph.
\item
The focus is on worst-case analysis of \emph{arbitrary graphs}.
Although flow-based methods achieve their worst-case $O(\log n)$ bounds on 
expanders, it is observed that \emph{spectral methods are appropriate for 
expander graphs} (basically, since the quadratic of a constant is a constant,
which implies a worst-case constant-factor approximation).
\item
Since the algorithmic task is simply to approximate the value of the 
objective function of Eqn.~(\ref{eqn:expansion_graph}) or 
Eqn.~(\ref{eqn:conductance_graph}), \emph{there is no particular interest in 
the actual pieces returned by the algorithm}, except insofar as those 
pieces have objective function value that is close to the optimum.%
\footnote{That is not to say that those pieces are not sometimes then used 
in a downstream data application.  Instead, it is to say simply that the 
algorithmic problem takes into account only the objective function value of 
those pieces and not how close those pieces might be to pieces achieving 
the optimum.}
\end{itemize}
Most of the TCS work is on the expansion versions of the graph partitioning 
problem, but it is noted that most of the results obtained ``go through'' to 
the conductance versions by considering appropriately-weighted functions on 
the nodes.

There has also been recent work within TCS that has been motivated by 
achieving improved worst-case bounds and/or being appropriate in data 
analysis applications where very large graphs, say with millions or 
billions or nodes, arise.
These methods include:
\begin{itemize}
\item
\emph{Local Spectral Methods}. 
These methods~\cite{Spielman:2004,andersen06local,chung07_fourproofs,Chung07_localcutsLAA,Chung07_heatkernelPNAS}
take as input a seed node and a locality parameter and return as output a 
``good'' cluster ``nearby'' the seed node. 
Moreover, they often have computational cost that is proportional to the 
size of the piece returned, and 
they have roughly the same kind of quadratic approximation guarantees as the 
global spectral method.
\item
\emph{Cut-Improvement Algorithms}.
These methods~\cite{Gallo:1989,kevin04mqi,andersen08soda,MOV09_TR}, 
\emph{e.g.}, MQI~\cite{kevin04mqi,andersen08soda}, take as input a graph 
and an initial cut and use spectral or flow-based methods to return as 
output a ``good'' cut that is ``within'' or ``nearby'' the original cut.
As such, they, can be combined with a good method (say, from spectral or 
Metis) for initially splitting the graph into two pieces to obtain a 
heuristic method for finding low conductance cuts in the 
whole graph~\cite{kevin04mqi,andersen08soda}. 
\item
\emph{Combining Spectral and Flow}.
These methods can be viewed as combinations of spectral and flow-based 
techniques which exploit the complementary strengths of these two classes of 
techniques.
They include an algorithm that used semidefinite programming to find a 
solution that is within a multiplicative factor of $O(\sqrt{\log n})$ of 
optimal~\cite{Arora:2004,ARV_CACM08}, as well as several related 
algorithms~\cite{AHK04,khandekar06_partitioning,Arora:2007,OSVV08,LMO09} 
that are more amenable to medium- and large-scale implementation.
\end{itemize}

\subsection{A novel approach to characterizing network structure}

Most work on community detection in large networks begins with an 
observation such as:
``It is a matter of common experience that communities exist in networks.
Although not precisely defined, communities are usually thought of as sets 
of nodes with better connections amongst its members than with the rest of 
the world.''
(I have not provided a reference for this quotation since variants of it 
could have been drawn from any one of scores or hundreds of papers on the 
topic.
Most of this work then typically goes on to apply the \emph{modus operandi} 
described previously.)
Although far from perfect, conductance is probably the combinatorial 
quantity that most closely captures this intuitive bi-criterial notion of 
what it means for a set of nodes to be a good community.

Even more important from the perspective of this chapter, the use of 
conductance as an objective function has the following benefit.
Although exactly solving the combinatorial problem of 
Eqn.~(\ref{eqn:expansion_graph}) or Eqn.~(\ref{eqn:conductance_graph}) is 
intractable, there exists a wide range of heuristics and approximation 
algorithms, the respective strengths and weaknesses of which are 
well-understood in theory and/or in practice, for approximately optimizing 
conductance. 
In particular, recall that spectral methods and multi-commodity flow-based 
methods are complementary in that:
\begin{itemize}
\item
The worst-case $O(\log n)$ approximation factor is obtained for flow-based 
methods on expander 
graphs~\cite{Leighton:1988,Leighton:1999}, a class of graphs which does not 
cause problems for spectral methods (to the extent than any methods are 
appropriate).
\item
The worst-case quadratic approximation factor is obtained for spectral 
methods on graphs with ``long stringy'' 
pieces~\cite{guatterymiller98,spielman96_spectral}, basically since spectral 
methods can confuse ``long path'' with ``deep cuts,'' a difference that does
not cause problems for flow-based methods. 
\item
Both methods perform well on ``low-dimensional graphs''---\emph{e.g.}, road 
networks, discretizations of low-dimensional spaces as might be encountered 
in scientific modeling, graphs that are easily-visualizable in two 
dimensions, and other graphs that have good well-balanced cuts---although 
the biases described in the previous two bullets still manifest themselves.
\end{itemize}
I should note that empirical 
evidence~\cite{LLDM08_communities_CONF,LLDM08_communities_TR,LLM10_communities_CONF} 
clearly demonstrates that large social and information networks have all of 
these properties---they are expander-like when viewed at large size scales; 
their sparsity and noise properties are such that they have structures 
analogous to stringy pieces that are cut off or regularized away by spectral 
methods; and they often have structural regions that at least locally are 
meaningfully low-dimensional.

All of this suggests a rather novel approach---that of using approximation 
algorithms for the intractable graph partitioning problem (or other 
intractable problems) as ``experimental probes'' of the structure of large 
social and information networks.
By this, I mean using these algorithms to ``cut up'' or ``tear apart'' a 
large network in order to provide insights into its structure.
From this perspective, although we are defining an edge-counting metric and 
then using it to perform an optimization, we are not particularly 
interested \emph{per se} in the clusters that are output by the algorithms.
Instead, we are interested in what these clusters, coupled with knowledge of 
the inner workings of the approximation algorithms, tell us about network 
structure.
Relatedly, given the noise properties and sparsity structure of large 
networks, we are not particularly interested \emph{per se} in the solution 
to the combinatorial problem.
For example, if we were provided with an oracle that returned the solution 
to the intractable combinatorial problem, it would be next to useless to us.
Instead, much more information is revealed by looking at ensembles of output
clusters in light of the artifactual properties of the heuristics and 
approximation algorithms employed.
That is, from this perspective, we are interested in the \emph{statistical 
properties implicit in worst-case approximation algorithms} and what these 
properties tell us about the data.

The analogy between approximation algorithms for the intractable graph 
partitioning problem and experimental probes is
meant to be taken more seriously rather than less.
Recall, \emph{e.g.}, that it is a non-trivial exercise to determine what a 
protein or DNA molecule ``looks like''---after all, such a molecule is very 
small, it can't easily be visualized with traditional methods, it has 
complicated dynamical properties, etc.
To determine the structure of a protein, an experimentalist puts the protein 
in solution or in a crystal and then probes it with X-ray crystallography or 
a nuclear magnetic resonance signal.
In more detail, one sends in a signal that scatters off the protein; one 
measures a large quantity of scattering output that comes out the other 
side; and then, using what is measured as well as knowledge of the physics 
of the input signal, one reconstructs the structure of the hard-to-visualize 
protein.
In an analogous manner, we can analyze the structure of a large social or 
information by tearing it apart in numerous ways with one of several local 
or global graph partitioning algorithms; measuring a large number of pieces
output by these procedures; and then, using what was measured as well as 
knowledge of the artifactual properties of the heuristics and approximation 
algorithms employed, reconstructing structural properties of the 
hard-to-visualize network.

By adopting this approach, one can hope to use approximation algorithms to 
test commonly-made data analysis assumptions, \emph{e.g.}, that the data 
meaningfully lie on a manifold or some other low-dimensional space, or that 
the data have structures that are meaningfully-interpretable as communities.

\subsection{Community-identification applications of this approach}

As a proof-of-principle application of this approach, let me return to the 
question of characterizing the small-scale and large-scale community 
structure in large social and information networks.
Given the conductance (or some other) community quality score, define the 
\emph{network community profile} (NCP) as the conductance value, as a 
function of size $k$, of the minimum conductance set of cardinality $k$ in 
the network:
$$
\Phi(k) = \min_{ S \subset V,|S|=k} \phi(S)  .
$$
Intuitively, just as the ``surface-area-to-volume'' aspect of conductance 
captures the ``gestalt'' notion of a good cluster or community, the NCP 
measures the score of ``best'' community as a function of community size in 
a network. 
See Figure~\ref{fig:communities:toy-fig2} for an illustration of small 
network and Figure~\ref{fig:communities:toy-fig2-ncp} for the corresponding 
NCP.
Operationally, since the NCP is NP-hard to compute exactly, one can use
approximation algorithms for solving the Minimum Conductance Cut Problem in 
order to compute different approximations to it.
Although we have experimented with a wide range of procedures on smaller 
graphs, in order to scale to very large social and information networks,%
\footnote{We have examined a large number of networks with these methods.  
The networks we studied range in size from tens of nodes and scores of 
edges up to millions of nodes and tens of millions of edges.  The networks 
were drawn from a wide range of domains and included large social networks, 
citation networks, collaboration networks, web graphs, communication 
networks, citation networks, internet networks, affiliation networks, and 
product co-purchasing networks.
See~\cite{LLDM08_communities_CONF,LLDM08_communities_TR,LLM10_communities_CONF} 
for details.}
we used:
\begin{itemize}
\item
\emph{Metis+MQI}---this flow-based method consists of using the popular 
graph partitioning package Metis~\cite{karypis98_metis} followed by a 
flow-based MQI cut-improvement post-processing step~\cite{kevin04mqi}, and 
it provides a surprisingly strong heuristic method for finding good 
conductance cuts;
\item
\emph{Local Spectral}---although this spectral method~\cite{andersen06local} 
is worse than Metis+MQI at finding very good conductance pieces, the pieces 
it finds tend to be ``tighter'' and more ``community-like'';~and 
\item
\emph{Bag-of-Whiskers}---this is a simple heuristic to paste together small 
barely-connected sets of nodes that exert a surprisingly large influence on 
the NCP of large informatics graphs;
\end{itemize}
and we compared and contrasted the outputs of these different procedures.

The NCP behaves in a characteristic downward-sloping 
manner~\cite{LLDM08_communities_CONF,LLDM08_communities_TR,LLM10_communities_CONF}
for graphs that are well-embeddable into an underlying low-dimensional 
geometric structure, \emph{e.g.}, low-dimensional lattices, road networks, 
random geometric graphs, and data sets that are well-approximatable by 
low-dimensional spaces and non-linear manifolds.
(See, \emph{e.g.}, Figure~\ref{fig:communities:toy-fig2-ncp}.)
Relatedly, a downward sloping NCP is also observed for small 
commonly-studied networks that have been widely-used as testbeds for 
community identification algorithms, while the NCP is roughly flat at all 
size scales for well-connected expander-like graphs such as moderately-dense 
random graphs. 
Perhaps surprisingly, common generative models, including preferential 
attachment models, as well as ``copying'' models and hierarchical models 
specifically designed to reproduce community structure, also have flat or 
downward-sloping NCPs.
Thus, when viewed from the perspective of $10,000$ meters, all of these 
graphs ``look like'' either a hot dog (in the sense that one can split the 
graph into two large and meaningful pieces) or a moderately-dense 
expander-like random graph (in the sense that there are no good partitions 
of any size in the graph).
This fact is used to great advantage by common machine learning and data 
analysis tools.

From the same perspective of $10,000$ meters, what large social and 
information networks ``look like'' is very different.
In Figure~\ref{fig:communities:realNCP-LiveJournal} and
Figure~\ref{fig:communities:realNCP-epinions}, I present the NCP computed 
several different ways for two representative large social and information 
networks.
Several things are worth observing:
first, up to a size scale of roughly $100$ nodes, the NCP goes down;
second, it then achieves its global minimum;
third, above that size scale, the NCP gradually increases, indicating that 
the community-quality of possible communities gets gradually worse and worse 
as larger and larger purported communities are considered; and
finally, even at the largest size scales there is substantial residual 
structure above and beyond what is present in a corresponding random graph.
Thus, good network communities, at least as traditionally conceptualized, 
tend to exist only up to a size scale of roughly $100$ nodes, while at 
larger size scales network communities become less and less community-like.

The explanation for this phenomenon is that large social and information 
networks have a \emph{nested core-periphery structure}, in which the network 
consists of a large moderately-well connected core and a large number of 
very well-connected communities barely connected to the core, as illustrated
in Figure~\ref{fig:communities:core-periphery}.
If we define a notion of a \emph{whisker} to be maximal sub-graph detached 
from network by removing a single edge and the associated notion of the
\emph{core} to be the rest of the graph, \emph{i.e.}, the 2-edge-connected 
core, then whiskers contain (on average) $40\%$ of nodes and $20\%$ of 
edges.
Moreover, in nearly every network, the global minimum of the NCP is a 
whisker.
Importantly, though, even if \emph{all} the whiskers are removed, then the 
core itself has a downward-then-upward-sloping NCP, indicating that the 
core itself has a nested core-periphery structure.
In addition to being of interest in community identification applications, 
this fact explains the inapplicability of many common machine learning and 
data analysis tools for large informatics graphs.
For example, it implies that recursive-bisection heuristics are largely
inappropriate for this class of graphs (basically since the recursion depth 
will be very large).

There are numerous ways one can be confident in these domain-specific 
conclusions.%
\footnote{For example, 
\emph{structural considerations} (\emph{e.g.}, small barely connected 
whiskers, discoverable with pretty much any heuristic or algorithm, are 
responsible for minimum);
\emph{lower-bound considerations} (\emph{e.g.}, SDP-based lower bounds for 
large partitions provide a strong nonexistence results for large 
communities); 
\emph{domain-specific considerations} (\emph{e.g.}, considering the 
properties of ``ground truth'' communities and the comparing results from 
networks in different application areas with very different degrees of 
homogeneity to their edge semantics); and
\emph{alternate-formalization considerations} (\emph{e.g.}, other 
formalizations of the community concept that take into account the 
inter-connectivity versus intra-connectivity bi-criterion tend to have 
similar NCPs, while formalizations that take into account one or the other
criteria behave very differently).}
Most relevant for the interplay between the algorithmic and statistical 
perspectives on data that I am discussing in this chapter are:
\begin{itemize}
\item
\emph{Modeling Considerations}.
Several lines of evidence point to the role that sparsity and noise have in 
determining the large-scale clustering structure of large networks.
\begin{itemize}
\item
Extremely sparse Erd\H{o}s-R\'{e}nyi random graphs with connection 
probability parameter roughly $p \in (1/n, \log (n)/n)$---\emph{i.e.}, in the 
parameter regime where measure fails to concentrate sufficiently to have a 
fully-connected graph or to have observed node degrees close to expected 
node degrees---provide the simplest mechanism to reproduce the qualitative 
property of having very-imbalanced deep cuts and no well-balanced deep cuts.
\item
So-called Power Law random 
graphs~\cite{Chung02_distancesPNAS,Chung03_spectraPNAS} with power law 
parameter roughly $\beta \in (2,3)$---which, due to exogenously-specified node 
degree variability, have an analogous failure of measure 
concentration---also exhibit this same qualitative property.
\item
A model in which new edges are added randomly but with an 
iterative ``forest fire'' burning mechanism provides a mechanism to 
reproduce the qualitative property of a relatively 
\emph{gradually-increasing} NCP.%
\footnote{As an aside, these considerations suggest that the data look like 
\emph{local structure on top of a global sparse quasi-random scaffolding}, 
much more than the more commonly-assumed model of \emph{local noise on top 
of global geometric or hierarchical structure}, an observation with 
significant modeling implications for this class of data.}
\end{itemize}
\item
\emph{Algorithmic-Statistical Considerations}.
Ensembles of clusters returned by different approximation algorithms, 
\emph{e.g.}, Local Spectral versus the flow-based methods such as Metis+MQI 
versus the Bag-of-Whiskers heuristic, have very different properties, in a 
manner consistent with known artifactual properties of how those algorithms 
operate. 
\begin{itemize}
\item
For example, Metis+MQI finds sets of nodes that have very good conductance 
scores---at very small size scales, these are similar to the pieces from 
Local Spectral and these sets of nodes could plausibly be interpreted as 
good communities; but at size scales larger than roughly $100$ nodes, these
are often tenuously-connected (and in some cases unions of disconnected) 
pieces, for which such an interpretation is tenuous at best.
\item
Similarly, the NCP of a variant of flow-based partitioning that permits 
disconnected pieces to be output mirrors the NCP from the Bag-of-Whiskers 
heuristic that combines disconnected whiskers, while the NCP of a variant 
of flow-based partitioning that requires connected pieces more closely 
follows that of Local Spectral, at least until much larger size scales where 
flow has problems since the graph becomes more expander-like.
\item
Finally, Local Spectral finds sets of nodes with good (but not as good as 
Metis+MQI) conductance value that that are more ``compact'' and thus more 
plausibly community-like than those pieces returned by Metis+MQI. 
For example, since spectral methods confuse long paths with deep cuts,
empirically one obtains sets of nodes that have worse conductance scores but 
are internally more coherent in that, \emph{e.g.}, they have substantially 
smaller diameter and average diameter.
This is illustrated in Figures~\ref{fig:communities:spectral-cmty-1} 
through~\ref{fig:communities:flow-cmty-2}:
Figures~\ref{fig:communities:spectral-cmty-1} 
and~\ref{fig:communities:spectral-cmty-2} show two ca. $500$-node 
communities obtained from Local Spectral, which are thus more internally 
coherent than the two ca. $500$-node communities illustrated in 
Figures~\ref{fig:communities:flow-cmty-1} 
and~\ref{fig:communities:flow-cmty-2} which were obtained from Metis+MQI and
which thus consist of two tenuously-connected smaller pieces. 
\end{itemize}
\end{itemize}

\subsection{Some general thoughts on statistical issues and graph algorithms}

I will conclude this section with a few general thoughts raised by these
empirical results.
Note that the modeling considerations suggest, and the 
algorithmic-statistical considerations verify, that statistical issues, and 
in particular \emph{regularization issues}, are particularly important for 
extracting domain-specific understanding from approximation algorithms 
applied this class of data.

By regularization, of course, I mean the traditional notion that the computed 
quantities---clusters and communities in this case---are empirically 
``nicer,'' or more ``smooth'' or ``regular,'' in some useful domain-specific 
sense~\cite{Neu98,CH02,BL06}.
Recall that regularization grew out of the desire to find meaningful 
solutions to ill-posed problems in integral equation theory, and it may be 
viewed as adding a small amount of bias to reduce substantially the variance 
or variability of a computed quantity.
It is typically implemented by adding a ``loss term'' to the objective 
function being optimized.
Although this manner of implementation leads to a natural interpretation of
regularization as a trade-off between optimizing the objective and avoiding
over-fitting the data, it typically leads to optimization problems that 
are harder (think of $\ell_1$-regularized $\ell_2$-regression) or at least
no easier (think of $\ell_2$-regularized $\ell_2$-regression) than the 
original problem, a situation that is clearly unacceptable in the very 
large-scale applications I have been discussing.

Interestingly, though, at least for very large and extremely sparse social 
and information networks, intuitive notions of cluster quality tend to fail 
as one aggressively optimizes the community-quality 
score~\cite{LLM10_communities_CONF}.
For instance, by aggressively optimizing conductance with Metis+MQI, one 
obtains disconnected or barely-connected clusters that do not correspond to 
intuitive communities. 
This suggests the rather interesting notion of implementing
\emph{implicit regularization via approximate computation}---approximate 
optimization of the community score in-and-of-itself introduces a 
systematic bias in the properties of the extracted clusters, when compared 
with much more aggressive optimization or computation of the intractable 
combinatorial optimum. 
Of course, such a bias may in fact be preferred since, as in case of Local 
Spectral, the resulting ``regularized communities'' are more compact and 
correspond to more closely to intuitive communities.
Explicitly incorporating the tradeoff between the conductance of the 
bounding cut of the cluster and the internal cluster compactness into a 
new modified objective function would not have led to a more tractable 
problem; but implicitly incorporating it in the approximation algorithm had 
both algorithmic and statistical benefits.
It is clearly of interest to formalize this approach more generally.

\section{Conclusions and Future Directions}

As noted above, the algorithmic and statistical perspectives on data and 
what might be interesting questions to ask of the data are quite different, 
but they are not incompatible.
Some of the more prominent recent examples of this include: 
\begin{itemize}
\item
Statistical, randomized, and probabilistic ideas are central to much of 
the recent work on developing improved worst-case approximation algorithms 
for matrix problems.
\item
Otherwise intractable optimization problems on graphs and networks yield to 
approximation algorithms when assumptions are made about the network 
participants. 
\item
Much recent work in machine learning draws on ideas from both areas; 
similarly, in combinatorial scientific computing, implicit knowledge about 
the system being studied often leads to good heuristics for combinatorial 
problems.
\item
In boosting, a statistical technique that fits an additive model by 
minimizing an objective function with a method such as gradient descent, the 
computation parameter, \emph{i.e.}, the number of iterations, also serves as 
a regularization parameter.
\end{itemize}
In this chapter, I have focused on two examples that illustrate this point 
in a somewhat different way.
For the first example, we have seen that using a concept fundamental to 
statistical and diagnostic regression analysis was crucial in the 
development of improved algorithms that come with worst-case performance 
guarantees and that, for related reasons, also perform well in practice.
For the second example, we have seen that an improved understanding of 
worst-case algorithms (\emph{e.g.}, when they perform well, when they 
perform poorly, and what regularization is implicitly implemented by them)
suggested methods for using those approximation algorithms as ``experimental 
probes'' of large informatics graphs and for better inference and prediction 
on those graphs.

To conclude, I should re-emphasize that in neither of these applications 
did we \emph{first} perform statistical modeling, independent of algorithmic 
considerations, and \emph{then} apply a computational procedure as a black 
box.
This approach of more closely coupling the computational procedures used 
with statistical modeling or statistical understanding of the data seems 
particularly appropriate more generally for very large-scale data analysis 
problems, where design decisions are often made based on computational 
constraints but where it is of interest to understand the implicit 
statistical consequences of those decisions.

\vspace{0.25in}
\noindent
\textbf{Acknowledgments}

\noindent
I would like to thank 
Schloss Dagstuhl and the organizers of Dagstuhl Seminar 09061 for an 
enjoyable and fruitful meeting; 
Peristera Paschou, Petros Drineas, and Christos Boutsidis for fruitful 
discussions on DNA SNPs, population genetics, and matrix algorithms; 
Jure Leskovec and Kevin Lang for fruitful discussions on 
social and information networks and graph algorithms;
and my co-organizers of as well as numerous participants of the MMDS meetings, 
at which many of the ideas described here were formed.


\begin{thebibliography}{100}

\bibitem{AC06}
N.~Ailon and B.~Chazelle.
\newblock Approximate nearest neighbors and the fast {J}ohnson-{L}indenstrauss
  transform.
\newblock In {\em Proceedings of the 38th Annual ACM Symposium on Theory of
  Computing}, pages 557--563, 2006.

\bibitem{Alter_SVD_00}
O.~Alter, P.O. Brown, and D.~Botstein.
\newblock Singular value decomposition for genome-wide expression data
  processing and modeling.
\newblock {\em Proc. Natl. Acad. Sci. USA}, 97(18):10101--10106, 2000.

\bibitem{andersen06local}
R.~Andersen, F.R.K. Chung, and K.~Lang.
\newblock Local graph partitioning using {PageRank} vectors.
\newblock In {\em FOCS '06: Proceedings of the 47th Annual IEEE Symposium on
  Foundations of Computer Science}, pages 475--486, 2006.

\bibitem{andersen08soda}
R.~Andersen and K.~Lang.
\newblock An algorithm for improving graph partitions.
\newblock In {\em SODA '08: Proceedings of the 19th ACM-SIAM Symposium on
  Discrete algorithms}, pages 651--660, 2008.

\bibitem{AHK04}
S.~Arora, E.~Hazan, and S.~Kale.
\newblock ${O}(\sqrt {\log n)}$ approximation to sparsest cut in
  $\tilde{O}(n^2)$ time.
\newblock In {\em FOCS '04: Proceedings of the 45th Annual Symposium on
  Foundations of Computer Science}, pages 238--247, 2004.

\bibitem{Arora:2007}
S.~Arora and S.~Kale.
\newblock A combinatorial, primal-dual approach to semidefinite programs.
\newblock In {\em STOC '07: Proceedings of the 39th annual ACM Symposium on
  Theory of Computing}, pages 227--236, 2007.

\bibitem{Arora:2004}
S.~Arora, S.~Rao, and U.~Vazirani.
\newblock Expander flows, geometric embeddings and graph partitioning.
\newblock In {\em STOC '04: Proceedings of the 36th annual ACM Symposium on
  Theory of Computing}, pages 222--231, 2004.

\bibitem{ARV_CACM08}
S.~Arora, S.~Rao, and U.~Vazirani.
\newblock Geometry, flows, and graph-partitioning algorithms.
\newblock {\em Communications of the ACM}, 51(10):96--105, 2008.

\bibitem{AMT09_DRAFT}
H.~Avron, P.~Maymounkov, and S.~Toledo.
\newblock Blendenpik: Supercharging {LAPACK}'s least-squares solver.
\newblock Manuscript. (2009).

\bibitem{BB06}
M.W. Berry and M.~Browne.
\newblock Email surveillance using non-negative matrix factorization.
\newblock {\em Computational and Mathematical Organization Theory},
  11(3):249--264, 2005.

\bibitem{BL06}
P.~Bickel and B.~Li.
\newblock Regularization in statistics.
\newblock {\em TEST}, 15(2):271--344, 2006.

\bibitem{BH91}
C.~H. Bischof and P.~C. Hansen.
\newblock Structure-preserving and rank-revealing {QR}-factorizations.
\newblock {\em SIAM Journal on Scientific and Statistical Computing},
  12(6):1332--1350, 1991.

\bibitem{BQ98b}
C.~H. Bischof and G.~Quintana-Ort{\'i}.
\newblock Algorithm 782: {C}odes for rank-revealing {QR} factorizations of
  dense matrices.
\newblock {\em ACM Transactions on Mathematical Software}, 24(2):254--257,
  1998.

\bibitem{BQ98a}
C.~H. Bischof and G.~Quintana-Ort{\'i}.
\newblock Computing rank-revealing {QR} factorizations of dense matrices.
\newblock {\em ACM Transactions on Mathematical Software}, 24(2):226--253,
  1998.

\bibitem{BMD08_CSSP_TR}
C.~Boutsidis, M.W. Mahoney, and P.~Drineas.
\newblock An improved approximation algorithm for the column subset selection
  problem.
\newblock Technical report.
\newblock Preprint: arXiv:0812.4293 (2008).

\bibitem{BMD08_CSSP_KDD}
C.~Boutsidis, M.W. Mahoney, and P.~Drineas.
\newblock Unsupervised feature selection for principal components analysis.
\newblock In {\em Proceedings of the 14th Annual ACM SIGKDD Conference}, pages
  61--69, 2008.

\bibitem{BMD09_kmeans_NIPS}
C.~Boutsidis, M.W. Mahoney, and P.~Drineas.
\newblock Unsupervised feature selection for the $k$-means clustering problem.
\newblock In {\em Annual Advances in Neural Information Processing Systems 22:
  Proceedings of the 2009 Conference}, 2009.

\bibitem{Bri01_all}
L.~Breiman.
\newblock Statistical modeling: The two cultures (with comments and a rejoinder
  by the author).
\newblock {\em Statistical Science}, 16(3):199--231, 2001.

\bibitem{BG65}
P.~Businger and G.H. Golub.
\newblock Linear least squares solutions by {H}ouseholder transformations.
\newblock {\em Numerische Mathematik}, 7:269--276, 1965.

\bibitem{CR08_TR}
E.J. Candes and B.~Recht.
\newblock Exact matrix completion via convex optimization.
\newblock Technical report.
\newblock Preprint: arXiv:0805.4471 (2008).

\bibitem{Cha87}
T.F. Chan.
\newblock Rank revealing {QR}~factorizations.
\newblock {\em Linear Algebra and Its Applications}, 88/89:67--82, 1987.

\bibitem{CH90}
T.F. Chan and P.C. Hansen.
\newblock Computing truncated singular value decomposition least squares
  solutions by rank revealing {QR}-factorizations.
\newblock {\em SIAM Journal on Scientific and Statistical Computing},
  11:519--530, 1990.

\bibitem{CH94}
T.F. Chan and P.C. Hansen.
\newblock Low-rank revealing {QR}~factorizations.
\newblock {\em Numerical Linear Algebra with Applications}, 1:33--44, 1994.

\bibitem{CI94}
S.~Chandrasekaran and I.~C.~F. Ipsen.
\newblock On rank-revealing factorizations.
\newblock {\em SIAM Journal on Matrix Analysis and Applications}, 15:592--622,
  1994.

\bibitem{CH86}
S.~Chatterjee and A.S. Hadi.
\newblock Influential observations, high leverage points, and outliers in
  linear regression.
\newblock {\em Statistical Science}, 1(3):379--393, 1986.

\bibitem{ChatterjeeHadi88}
S.~Chatterjee and A.S. Hadi.
\newblock {\em Sensitivity Analysis in Linear Regression}.
\newblock John Wiley \& Sons, New York, 1988.

\bibitem{ChatterjeeHadiPrice00}
S.~Chatterjee, A.S. Hadi, and B.~Price.
\newblock {\em Regression Analysis by Example}.
\newblock John Wiley \& Sons, New York, 2000.

\bibitem{Cheeger69_bound}
J.~Cheeger.
\newblock A lower bound for the smallest eigenvalue of the laplacian.
\newblock In {\em Problems in Analysis, Papers dedicated to Salomon Bochner},
  pages 195--199. Princeton University Press, 1969.

\bibitem{CH02}
Z.~Chen and S.~Haykin.
\newblock On different facets of regularization theory.
\newblock {\em Neural Computation}, 14(12):2791--2846, 2002.

\bibitem{Chung:1997}
F.R.K. Chung.
\newblock {\em Spectral graph theory}, volume~92 of {\em CBMS Regional
  Conference Series in Mathematics}.
\newblock American Mathematical Society, 1997.

\bibitem{chung07_fourproofs}
F.R.K Chung.
\newblock Four proofs of {C}heeger inequality and graph partition algorithms.
\newblock In {\em Proceedings of ICCM}, 2007.

\bibitem{Chung07_heatkernelPNAS}
F.R.K. Chung.
\newblock The heat kernel as the pagerank of a graph.
\newblock {\em Proceedings of the National Academy of Sciences of the United
  States of America}, 104(50):19735--19740, 2007.

\bibitem{Chung07_localcutsLAA}
F.R.K. Chung.
\newblock Random walks and local cuts in graphs.
\newblock {\em Linear Algebra and its Applications}, 423:22--32, 2007.

\bibitem{Chung02_distancesPNAS}
F.R.K. Chung and L.~Lu.
\newblock The average distances in random graphs with given expected degrees.
\newblock {\em Proceedings of the National Academy of Sciences of the United
  States of America}, 99(25):15879--15882, 2002.

\bibitem{Chung03_spectraPNAS}
F.R.K. Chung, L.~Lu, and V.~Vu.
\newblock The spectra of random graphs with given expected degrees.
\newblock {\em Proceedings of the National Academy of Sciences of the United
  States of America}, 100(11):6313--6318, 2003.

\bibitem{CM09b}
A.~Civril and M.~Magdon-Ismail.
\newblock Column based matrix reconstruction via greedy approximation of {SVD}.
\newblock {\em Numerical Linear Algebra with Applications}, 000:000--000, 2009.

\bibitem{CM09a}
A.~Civril and M.~Magdon-Ismail.
\newblock On selecting a maximum volume sub-matrix of a matrix and related
  problems.
\newblock {\em Theoretical Computer Science}, 410:4801--4811, 2009.

\bibitem{DDLFH90}
S.T. Deerwester, S.T. Dumais, G.W. Furnas, T.K. Landauer, and R.~Harshman.
\newblock Indexing by latent semantic analysis.
\newblock {\em Journal of the American Society for Information Science},
  41(6):391--407, 1990.

\bibitem{DV06}
A.~Deshpande and S.~Vempala.
\newblock Adaptive sampling and fast low-rank matrix approximation.
\newblock In {\em Proceedings of the 10th International Workshop on
  Randomization and Computation}, pages 292--303, 2006.

\bibitem{dhillon07graclus}
I.S. Dhillon, Y.~Guan, and B.~Kulis.
\newblock Weighted graph cuts without eigenvectors: A multilevel approach.
\newblock {\em IEEE Transactions on Pattern Analysis and Machine Intelligence},
  29(11):1944--1957, 2007.

\bibitem{Donath:1972}
W.E. Donath and A.J. Hoffman.
\newblock Algorithms for partitioning graphs and computer logic based on
  eigenvectors of connection matrices.
\newblock {\em IBM Technical Disclosure Bulletin}, 15(3):938--944, 1972.

\bibitem{Donoho00}
D.L. Donoho.
\newblock Aide-memoire. {H}igh-dimensional data analysis: The curses and
  blessings of dimensionality, 2000.
\newblock
  \texttt{http://www-stat.stanford.edu/~donoho/Lectures/AMS2000/Curses.pdf}
  Accessed December, 17 2008.

\bibitem{DS66}
N.~R. Draper and D.~M. Stoneman.
\newblock Testing for the inclusion of variables in linear regression by a
  randomisation technique.
\newblock {\em Technometrics}, 8(4):695--699, 1966.

\bibitem{DFKVV04_JRNL}
P.~Drineas, A.~Frieze, R.~Kannan, S.~Vempala, and V.~Vinay.
\newblock Clustering large graphs via the singular value decomposition.
\newblock {\em Machine Learning}, 56(1-3):9--33, 2004.

\bibitem{dkm_matrix1}
P.~Drineas, R.~Kannan, and M.W. Mahoney.
\newblock Fast {Monte Carlo} algorithms for matrices {I}: Approximating matrix
  multiplication.
\newblock {\em SIAM Journal on Computing}, 36:132--157, 2006.

\bibitem{dkm_matrix2}
P.~Drineas, R.~Kannan, and M.W. Mahoney.
\newblock Fast {Monte Carlo} algorithms for matrices {II}: Computing a low-rank
  approximation to a matrix.
\newblock {\em SIAM Journal on Computing}, 36:158--183, 2006.

\bibitem{dkm_matrix3}
P.~Drineas, R.~Kannan, and M.W. Mahoney.
\newblock Fast {Monte Carlo} algorithms for matrices {III}: Computing a
  compressed approximate matrix decomposition.
\newblock {\em SIAM Journal on Computing}, 36:184--206, 2006.

\bibitem{DMM06}
P.~Drineas, M.W. Mahoney, and S.~Muthukrishnan.
\newblock Sampling algorithms for $\ell_2$ regression and applications.
\newblock In {\em Proceedings of the 17th Annual ACM-SIAM Symposium on Discrete
  Algorithms}, pages 1127--1136, 2006.

\bibitem{DMM08_CURtheory_JRNL}
P.~Drineas, M.W. Mahoney, and S.~Muthukrishnan.
\newblock Relative-error {CUR} matrix decompositions.
\newblock {\em SIAM Journal on Matrix Analysis and Applications}, 30:844--881,
  2008.

\bibitem{DMMS07_FastL2_TR}
P.~Drineas, M.W. Mahoney, S.~Muthukrishnan, and T.~Sarl\'{o}s.
\newblock Faster least squares approximation.
\newblock Technical report.
\newblock Preprint: arXiv:0710.1435 (2007).

\bibitem{FSS96}
U.~Fayyad, G.~Piatetsky-Shapiro, and P.~Smyth.
\newblock From data mining to knowledge discovery in databases.
\newblock {\em AI Magazine}, 17:37--54, 1996.

\bibitem{Fiduccia:1982}
C.M. Fiduccia and R.M. Mattheyses.
\newblock A linear-time heuristic for improving network partitions.
\newblock In {\em DAC '82: Proceedings of the 19th ACM/IEEE Conference on
  Design Automation}, pages 175--181, 1982.

\bibitem{fiedler73graphs}
M.~Fiedler.
\newblock Algebraic connectivity of graphs.
\newblock {\em Czechoslovak Mathematical Journal}, 23(98):298--305, 1973.

\bibitem{Fos86}
L.~V. Foster.
\newblock Rank and null space calculations using matrix decomposition without
  column interchanges.
\newblock {\em Linear Algebra and Its Applications}, 74:47--71, 1986.

\bibitem{gaertler05_clustering}
M.~Gaertler.
\newblock Clustering.
\newblock In U.~Brandes and T.~Erlebach, editors, {\em Network Analysis:
  Methodological Foundations}, pages 178--215. Springer, 2005.

\bibitem{Gallo:1989}
G.~Gallo, M.D. Grigoriadis, and R.E. Tarjan.
\newblock A fast parametric maximum flow algorithm and applications.
\newblock {\em SIAM Journal on Computing}, 18(1):30--55, 1989.

\bibitem{MMDS06}
G.~H. Golub, M.~W. Mahoney, P.~Drineas, and L.-H. Lim.
\newblock Bridging the gap between numerical linear algebra, theoretical
  computer science, and data applications.
\newblock {\em SIAM News}, 39(8), October 2006.

\bibitem{GVL96}
G.H. Golub and C.F.~Van Loan.
\newblock {\em Matrix Computations}.
\newblock Johns Hopkins University Press, Baltimore, 1996.

\bibitem{Gould96}
S.J. Gould.
\newblock {\em The Mismeasure of Man}.
\newblock W. W. Norton and Company, New York, 1996.

\bibitem{GE96}
M.~Gu and S.C. Eisenstat.
\newblock Efficient algorithms for computing a strong rank-revealing {QR}
  factorization.
\newblock {\em SIAM Journal on Scientific Computing}, 17:848--869, 1996.

\bibitem{guatterymiller98}
S.~Guattery and G.L. Miller.
\newblock On the quality of spectral separators.
\newblock {\em SIAM Journal on Matrix Analysis and Applications}, 19:701--719,
  1998.

\bibitem{HW78}
D.C. Hoaglin and R.E. Welsch.
\newblock The hat matrix in regression and {ANOVA}.
\newblock {\em The American Statistician}, 32(1):17--22, 1978.

\bibitem{HP92}
Y.~P. Hong and C.~T. Pan.
\newblock Rank-revealing {QR}~factorizations and the singular value
  decomposition.
\newblock {\em Mathematics of Computation}, 58:213--232, 1992.

\bibitem{Horne04}
B.D. Horne and N.J. Camp.
\newblock Principal component analysis for selection of optimal {SNP}-sets that
  capture intragenic genetic variation.
\newblock {\em Genetic Epidemiology}, 26(1):11--21, 2004.

\bibitem{jain99data}
A.K. Jain, M.N. Murty, and P.J. Flynn.
\newblock Data clustering: a review.
\newblock {\em ACM Computing Surveys}, 31:264--323, 1999.

\bibitem{KVV04_JRNL}
R.~Kannan, S.~Vempala, and A.~Vetta.
\newblock On clusterings: Good, bad and spectral.
\newblock {\em Journal of the ACM}, 51(3):497--515, 2004.

\bibitem{karypis98_metis}
G.~Karypis and V.~Kumar.
\newblock A fast and high quality multilevel scheme for partitioning irregular
  graphs.
\newblock {\em SIAM Journal on Scientific Computing}, 20:359--392, 1998.

\bibitem{karypis98metis}
G.~Karypis and V.~Kumar.
\newblock Multilevel k-way partitioning scheme for irregular graphs.
\newblock {\em Journal of Parallel and Distributed Computing}, 48:96--129,
  1998.

\bibitem{Kernighan:1970}
B.~Kernighan and S.~Lin.
\newblock An effective heuristic procedure for partitioning graphs.
\newblock {\em The Bell System Technical Journal}, pages 291--308, 1970.

\bibitem{khandekar06_partitioning}
R.~Khandekar, S.~Rao, and U.~Vazirani.
\newblock Graph partitioning using single commodity flows.
\newblock In {\em STOC '06: Proceedings of the 38th annual ACM Symposium on
  Theory of Computing}, pages 385--390, 2006.

\bibitem{KPS02}
F.G. Kuruvilla, P.J. Park, and S.L. Schreiber.
\newblock Vector algebra in the analysis of genome-wide expression data.
\newblock {\em Genome Biology}, 3:research0011.1--0011.11, 2002.

\bibitem{Lam03}
D.~Lambert.
\newblock What use is statistics for massive data?
\newblock In J.~E. Kolassa and D.~Oakes, editors, {\em Crossing boundaries:
  statistical essays in honor of Jack Hall}, pages 217--228. Institute of
  Mathematical Statistics, 2003.

\bibitem{LMO09}
K.~Lang, M.~W. Mahoney, and L.~Orecchia.
\newblock Empirical evaluation of graph partitioning using spectral embeddings
  and flow.
\newblock In {\em Proc. 8-th International SEA}, pages 197--208, 2009.

\bibitem{kevin04mqi}
K.~Lang and S.~Rao.
\newblock A flow-based method for improving the expansion or conductance of
  graph cuts.
\newblock In {\em IPCO '04: Proceedings of the 10th International IPCO
  Conference on Integer Programming and Combinatorial Optimization}, pages
  325--337, 2004.

\bibitem{Leighton:1988}
T.~Leighton and S.~Rao.
\newblock An approximate max-flow min-cut theorem for uniform multicommodity
  flow problems with applications to approximation algorithms.
\newblock In {\em FOCS '88: Proceedings of the 28th Annual Symposium on
  Foundations of Computer Science}, pages 422--431, 1988.

\bibitem{Leighton:1999}
T.~Leighton and S.~Rao.
\newblock Multicommodity max-flow min-cut theorems and their use in designing
  approximation algorithms.
\newblock {\em Journal of the ACM}, 46(6):787--832, 1999.

\bibitem{LLDM08_communities_TR}
J.~Leskovec, K.J. Lang, A.~Dasgupta, and M.W. Mahoney.
\newblock Community structure in large networks: Natural cluster sizes and the
  absence of large well-defined clusters.
\newblock arXiv:0810.1355, October 2008.

\bibitem{LLDM08_communities_CONF}
J.~Leskovec, K.J. Lang, A.~Dasgupta, and M.W. Mahoney.
\newblock Statistical properties of community structure in large social and
  information networks.
\newblock In {\em WWW '08: Proceedings of the 17th International Conference on
  World Wide Web}, pages 695--704, 2008.

\bibitem{LLM10_communities_CONF}
J.~Leskovec, K.J. Lang, and M.W. Mahoney.
\newblock Empirical comparison of algorithms for network community detection.
\newblock In {\em WWW '10: Proceedings of the 19th International Conference on
  World Wide Web}, pages 631--640, 2010.

\bibitem{LA04}
Z.~Lin and R.B. Altman.
\newblock Finding haplotype tagging {SNPs} by use of principal components
  analysis.
\newblock {\em American Journal of Human Genetics}, 75:850--861, 2004.

\bibitem{LLR95_JRNL}
N.~Linial, E.~London, and Y.~Rabinovich.
\newblock The geometry of graphs and some of its algorithmic applications.
\newblock {\em Combinatorica}, 15(2):215--245, 1995.

\bibitem{Malik10_TR}
M.~Magdon-Ismail.
\newblock Row sampling for matrix algorithms via a non-commutative {B}ernstein
  bound.
\newblock Technical report.
\newblock Preprint: arXiv:1008.0587 (2010).

\bibitem{MMDS08_SiamNews}
M.~W. Mahoney, L.-H. Lim, and G.~E. Carlsson.
\newblock {MMDS} 2008: Algorithmic and statistical challenges in modern
  large-scale data analysis.
\newblock {\em SIAM News}, 42(1 \& 2), January/February and March 2009.

\bibitem{MOV09_TR}
M.~W. Mahoney, L.~Orecchia, and N.~K. Vishnoi.
\newblock A spectral algorithm for improving graph partitions.
\newblock Technical report.
\newblock Preprint: arXiv:0912.0681 (2009).

\bibitem{CUR_PNAS}
M.W. Mahoney and P.~Drineas.
\newblock {CUR} matrix decompositions for improved data analysis.
\newblock {\em Proc. Natl. Acad. Sci. USA}, 106:697--702, 2009.

\bibitem{Matousek08_RSA}
J.~Matou\v{s}ek.
\newblock On variants of the {J}ohnson--{L}indenstrauss lemma.
\newblock {\em Random Structures and Algorithms}, 33(2):142--156, 2008.

\bibitem{Meng03}
Z.~Meng, D.V. Zaykin, C.F. Xu, M.~Wagner, and M.G. Ehm.
\newblock Selection of genetic markers for association analyses, using linkage
  disequilibrium and haplotypes.
\newblock {\em American Journal of Human Genetics}, 73(1):115--130, 2003.

\bibitem{mohar91_survey}
B.~Mohar.
\newblock The {Laplacian} spectrum of graphs.
\newblock In Y.~Alavi, G.~Chartrand, O.R. Oellermann, and A.J. Schwenk,
  editors, {\em Graph Theory, Combinatorics, and Applications, Vol. 2}, pages
  871--898. Wiley, 1991.

\bibitem{Neu98}
A.~Neumaier.
\newblock Solving ill-conditioned and singular linear systems: A tutorial on
  regularization.
\newblock {\em SIAM Review}, 40:636--666, 1998.

\bibitem{newman05_betweenness}
M.E.J. Newman.
\newblock A measure of betweenness centrality based on random walks.
\newblock {\em Social Networks}, 27:39--54, 2005.

\bibitem{newman2006finding}
M.E.J. Newman.
\newblock Finding community structure in networks using the eigenvectors of
  matrices.
\newblock {\em Physical Review E}, 74:036104, 2006.

\bibitem{NJW01_spectral}
A.Y. Ng, M.I. Jordan, and Y.~Weiss.
\newblock On spectral clustering: Analysis and an algorithm.
\newblock In {\em NIPS '01: Proceedings of the 15th Annual Conference on
  Advances in Neural Information Processing Systems}, 2001.

\bibitem{OSVV08}
L.~Orecchia, L.~Schulman, U.V. Vazirani, and N.K. Vishnoi.
\newblock On partitioning graphs via single commodity flows.
\newblock In {\em Proceedings of the 40th Annual ACM Symposium on Theory of
  Computing}, pages 461--470, 2008.

\bibitem{Pan00}
C.-T. Pan.
\newblock On the existence and computation of rank-revealing {LU}
  factorizations.
\newblock {\em Linear Algebra and Its Applications}, 316:199--222, 2000.

\bibitem{PT99}
C.~T. Pan and P.~T.~P. Tang.
\newblock Bounds on singular values revealed by {QR}~factorizations.
\newblock {\em BIT Numerical Mathematics}, 39:740--756, 1999.

\bibitem{Paschou07a}
P.~Paschou, M.~W. Mahoney, A.~Javed, J.~R. Kidd, A.~J. Pakstis, S.~Gu, K.~K.
  Kidd, and P.~Drineas.
\newblock Intra- and interpopulation genotype reconstruction from tagging
  {SNP}s.
\newblock {\em Genome Research}, 17(1):96--107, 2007.

\bibitem{Paschou07b}
P.~Paschou, E.~Ziv, E.G. Burchard, S.~Choudhry, W.~Rodriguez-Cintron, M.W.
  Mahoney, and P.~Drineas.
\newblock {PCA}-correlated {SNP}s for structure identification in worldwide
  human populations.
\newblock {\em PLoS Genetics}, 3:1672--1686, 2007.

\bibitem{PS03}
T.~Poggio and S.~Smale.
\newblock The mathematics of learning: Dealing with data.
\newblock {\em Notices of the AMS}, 50(5):537--544, May 2003.

\bibitem{Pot96}
A.~Pothen.
\newblock Graph partitioning algorithms with applications to scientific
  computing.
\newblock In D.~E. Keyes, A.~H. Sameh, and V.~Venkatakrishnan, editors, {\em
  Parallel Numerical Algorithms}. Kluwer Academic Press, 1996.

\bibitem{RT08}
V.~Rokhlin and M.~Tygert.
\newblock A fast randomized algorithm for overdetermined linear least-squares
  regression.
\newblock {\em Proc. Natl. Acad. Sci. USA}, 105(36):13212--13217, 2008.

\bibitem{RS00}
S.T. Roweis and L.K. Saul.
\newblock Nonlinear dimensionality reduction by local linear embedding.
\newblock {\em Science}, 290:2323--2326, 2000.

\bibitem{RV07}
M.~Rudelson and R.~Vershynin.
\newblock Sampling from large matrices: an approach through geometric
  functional analysis.
\newblock {\em Journal of the ACM}, 54(4):Article 21, 2007.

\bibitem{Sarlos06}
T.~Sarl\'{o}s.
\newblock Improved approximation algorithms for large matrices via random
  projections.
\newblock In {\em Proceedings of the 47th Annual IEEE Symposium on Foundations
  of Computer Science}, pages 143--152, 2006.

\bibitem{SWHSL06}
L.~K. Saul, K.~Q. Weinberger, J.~H. Ham, F.~Sha, and D.~D. Lee.
\newblock Spectral methods for dimensionality reduction.
\newblock In O.~Chapelle, B.~Schoelkopf, and A.~Zien, editors, {\em
  Semisupervised Learning}, pages 293--308. MIT Press, 2006.

\bibitem{Schaeffer07_survey}
S.E. Schaeffer.
\newblock Graph clustering.
\newblock {\em Computer Science Review}, 1(1):27--64, 2007.

\bibitem{ShiMalik00_NCut}
J.~Shi and J.~Malik.
\newblock Normalized cuts and image segmentation.
\newblock {\em IEEE Transcations of Pattern Analysis and Machine Intelligence},
  22(8):888--905, 2000.

\bibitem{Shm96}
D.~B. Shmoys.
\newblock Cut problems and their application to divide-and-conquer.
\newblock In D.S. Hochbaum, editor, {\em Approximation Algorithms for {NP}-Hard
  Problems}, pages 192--235. PWS Publishing, 1996.

\bibitem{Smy00}
P.~Smyth.
\newblock Data mining: data analysis on a grand scale?
\newblock {\em Statistical Methods in Medical Research}, 9(4):309--327, 2000.

\bibitem{SS08a_STOC}
D.A. Spielman and N.~Srivastava.
\newblock Graph sparsification by effective resistances.
\newblock In {\em Proceedings of the 40th Annual ACM Symposium on Theory of
  Computing}, pages 563--568, 2008.

\bibitem{spielman96_spectral}
D.A. Spielman and S.-H. Teng.
\newblock Spectral partitioning works: Planar graphs and finite element meshes.
\newblock In {\em FOCS '96: Proceedings of the 37th Annual IEEE Symposium on
  Foundations of Computer Science}, pages 96--107, 1996.

\bibitem{Spielman:2004}
D.A. Spielman and S.-H. Teng.
\newblock Nearly-linear time algorithms for graph partitioning, graph
  sparsification, and solving linear systems.
\newblock In {\em STOC '04: Proceedings of the 36th annual ACM Symposium on
  Theory of Computing}, pages 81--90, 2004.

\bibitem{TSL00}
J.B. Tenenbaum, V.~de~Silva, and J.C. Langford.
\newblock A global geometric framework for nonlinear dimensionality reduction.
\newblock {\em Science}, 290:2319--2323, 2000.

\bibitem{VW81}
P.F. Velleman and R.E. Welsch.
\newblock Efficient computing of regression diagnostics.
\newblock {\em The American Statistician}, 35(4):234--242, 1981.

\bibitem{luxburg05_survey}
U.~von Luxburg.
\newblock A tutorial on spectral clustering.
\newblock Technical Report 149, Max Plank Institute for Biological Cybernetics,
  August 2006.

\bibitem{weiss99_segmentation}
Y.~Weiss.
\newblock Segmentation using eigenvectors: a unifying view.
\newblock In {\em ICCV '99: Proceedings of the 7th IEEE International
  Conference on Computer Vision}, pages 975--982, 1999.

\bibitem{zachary77karate}
W.W. Zachary.
\newblock An information flow model for conflict and fission in small groups.
\newblock {\em Journal of Anthropological Research}, 33:452--473, 1977.

\bibitem{zhao04cluto}
Y.~Zhao and G.~Karypis.
\newblock Empirical and theoretical comparisons of selected criterion functions
  for document clustering.
\newblock {\em Machine Learning}, 55:311--331, 2004.

\end{thebibliography}

\end{document}